\documentclass{aa}

\usepackage{graphicx}
\usepackage{txfonts}
\newcommand{\lw}[1]{\smash{\lower2.ex\hbox{#1}}}
\def\kms{km~s$^{-1}$}

\def\hii{{\rm H}{\scriptsize{\rm II}}}
\def\HII{{\rm H}{\scriptsize{\rm II}}}

\def\kms{\mbox{km~s$^{-1}$}}

\def\cmq{cm$^{-2}$}

\def\um{$\mu$m}
\def\deg{$^{\circ}$}
\def\Msun{\mbox{$M_\odot$}}
\def\Lsun{\mbox{$L_\odot$}}

\def\Vlsr{$V_{\rm LSR}$}
\def\Vsys{$V_{\rm sys}$}

\def\Snu{$S_{\nu}$}

\def\wat{H$_2$O}
\def\amm{NH$_3$}

\def\HCO{\mbox{HCO$^+$}}
\def\hcop{\mbox{HCO$^+$}}

\def\mcn{\mbox{CH$_3$CN}}
\def\mcniso{\mbox{CH$_3^{13}$CN}}

\def\MCNII{C${{\rm H}_3}^{13}$CN}

\def\tCO{$^{13}$CO}
\def\CeO{C$^{18}$O}
\def\HtCOp{H$^{13}$CO$^+$}

\def\pbeam{beam$^{-1}$}

\def\sgm{\mbox{$\sigma$}}
\def\gs{\mbox{$\sigma$}}

\def\Mdust{\mbox{$M_{\rm dust}$}}

\def\td{\mbox{$t_{\rm d}$}}

\def\Vt{\mbox{$V_{\rm t}$}}
\def\Vb{\mbox{$V_{\rm b}$}}

\def\Mlobe{\mbox{$M_{\rm lobe}$}}

\def\Trot{\mbox{$T_{\rm rot}$}}
\def\Te{\mbox{$T_{\rm e}$}}
\def\Tmb{\mbox{$T_{\rm mb}$}}
\def\Tsys{\mbox{$T_{\rm sys}$}}
\def\Tmb{\mbox{$T_{\rm mb}$}}

\def\Tsb{\mbox{$T_{\rm sb}$}}

\def\Tex{\mbox{$T_{\rm ex}$}}

\def\Tdust{\mbox{$T_{\rm dust}$}}

\def\Rs{$R_{\rm s}$}
\def\Nly{$N_{\rm Ly}$}
\def\gn{G\,19.61$-$0.23}
\def\lesssim{\mathrel{\hbox{\rlap{\hbox{\lower4pt\hbox{$\sim$}}}\hbox{$<$}}}}
\def\gtrsim{\mathrel{\hbox{\rlap{\hbox{\lower4pt\hbox{$\sim$}}}\hbox{$>$}}}}

\def\thetahpbw{$\theta_{\rm HPBW}$}

\begin{document}
   \title{Infall, outflow, and rotation in the G19.61-0.23 hot molecular core}


   \author{R. S. Furuya\inst{1},
        R. Cesaroni\inst{2},          
        \and
        H. Shinnaga\inst{3}
      }

   \institute{Subaru Telescope, National Astronomical Observatory of Japan,
              650 North Aoh\'oku Place, Hilo, HI\,96720, U.S.A.,
              \email{rsf@subaru.naoj.org} 
        \and
             INAF-Osservatorio Astrofisico di Arcetri, Largo Enrico Fermi 5, I-50125 Firenze, Italy, 
             \email{cesa@arcetri.astro.it}
        \and
             Caltech Submillimeter Observatory, California Institute of Technology, 
111 Nowelo Street, Hilo, HI\,96720, U.S.A.,
             \email{shinnaga@submm.caltech.edu}
           \thanks{Preprint with full-resolution figures is available at \texttt{http://subarutelescope.org/staff/rsf/publication.html}}
           }

  \date{Received 2010 March 15; accepted 2010 June 20}

  \abstract
{}
   {The main goal of this study is to perform a sub-arcsecond resolution
analysis
of the high-mass star formation region \gn, both in the continuum and
molecular line emission. While the centimeter continuum images will be
discussed in detail in a forthcoming paper, here we focus on the (sub)mm
emission, devoting special attention to the hot molecular core.
}
%
   {A set of multi-wavelength continuum and molecular line emission data 
between 6\,cm and 890 \um\ were taken with the Very Large Array (VLA),
Nobeyama Millimeter Array (NMA),
Owens Valley Radio Observatory (OVRO), and
Submillimeter Array (SMA).
These data were analyzed in conjuction with previously published data.
}
   {Our observations resolve the HMC into three cores
whose masses are on the order of $10^1-10^3$~\Msun.
No submm core presents detectable free-free emission in the centimeter regime,
but they appear to be associated with masers and thermal line emission
from complex organic molecules.
Towards the most massive core, SMA1, the \mcn\ ($18_K-17_K$) lines
reveal hints of rotation
about the axis of a jet/outflow traced by \wat\ maser and \HtCOp(1--0) line emission.
Inverse P-Cygni profiles of the \tCO\ (3--2) and \CeO\ (3--2) lines seen towards
SMA1 indicate that the
central high-mass (proto)star(s) is (are) still gaining
mass with an accretion rate
$\ge 3 ~10^{-3}$ \Msun\ yr$^{-1}$.
Due to the linear scales and the large values of the accretion rate, we
hypothesize that we are observing an
accretion flow towards a cluster in the making,
rather than towards a single massive star.
}
{}

   \keywords{ISM: evolution --- HII regions ---  individual (G19.61-0.23)  --- jets and outflows --- Stars: early-type --- Submillimeter: ISM
               }
   \maketitle
%

\section{Introduction}
\label{s:i}

Understanding the formation of high-mass ($M_{\ast}\gtrsim 8$\Msun ) stars, 
as well as their evolution,
is the key to investigate the origin of the diversity of stars and stellar clusters in the Galaxy.
This is because high-mass stars not only affect the evolution of the interstellar 
medium in general, but are also preferentially born in clusters containing
a large number of low-mass stars, 
like, e.g., in the Trapezium cluster in Orion (\cite{muench02}).
Despite their importance, the formation process and early evolution of OB stars is
still far from being understood.
In the past decade, observations have suggested the existence of
flattened rotating disk-like structures around newly formed massive (proto)stars.
In particular, Keplerian disks appear to exist around B-type stars, whereas
only massive toroids have been found in association with
O-type stars. These toroids are very different from circumstellar disks, as they
appear to be transient, non-equilibrium entities (\cite{cesa07}). 
In spite of the substantial difference between the two types of objects,
their existence suggests that high-mass star formation may also
proceed through accretion with angular momentum conservation, as well as
in the case of low-mass stars.\par

This scenario implies the existence of infalling gas onto the newly formed
stars. Detection of gas infall in the vicinity of high-mass stars is
hampered by confusion with other processes such as outflow and rotation.
Nonetheless, to date interferometric observations 
at centimeter and (sub)millimeter wavelengths have succeeded 
in detecting infall in a handful of cases
(e.g., \cite{keto88}; \cite{cesa92}; \cite{maite06}; \cite{zapata08}; \cite{girart09})
lending support to the accretion scenario.
All these findings, however, do not suffice to explain the clustered mode of
OB-type star formation, which is another crucial issue in high-mass star
formation theories. A way to shed light on this process is to perform
detailed observational studies of selected star forming regions containing a
large number of very young OB-type (proto)stars, possibly in different
evolutionary phases. Excellent signposts of these are ultra compact (UC)
\hii\ regions and hot molecular cores (HMCs) (see e.g. \cite{kurtz00}).\par

%
\begin{table*}
\begin{minipage}[ht]{\columnwidth}
\caption{Summary of Interferometric Continuum Emission Imaging}
\label{tbl:contobs}
\centering
\renewcommand{\footnoterule}{}  
\begin{tabular}{rccccccc}
\hline \hline
\lw{Frequency} & \lw{Array} & Spatial Frequency  & 
\lw{$\theta_{\rm LAS}$\footnote{The largest detectable size scale.}} & \multicolumn{2}{c}{Synthesized Beam} & Image & \lw{Note} \\
\cline{5-6}
          &           & Range        &          & $\theta_{\rm major}\times\theta_{\rm minor}$ & P.A. & Noise Level & \\
(GHz)     &           & (k$\lambda$) & (arcsec) & (arcsec) & (deg) & (mJy beam$^{-1}$)  & \\
\hline
  4.860 & VLA-A, C        & 0.424 -- 596.7 & 485   & 0.630\,$\times$\,0.470 &   1.9   & 0.78 & this work \\
  8.415 & VLA-A, B, C, D  & 0.78 -- 1027.8 & 264   & 0.329\,$\times$\,0.228 &  24.4   & 18.1 & paper I \\
 14.940 & VLA-C, B, CnB   & 7.72 -- 554.0  &  26.7 & 0.550\,$\times$\,0.500 & $-8.4$  & 1.26 & paper I \\
 22.272 & VLA-B, D        & 1.96 -- 826.8  & 105   & 0.357\,$\times$\,0.260 & $-4.3$  & 0.47 & this work  \\
 43.340 & VLA-D           & 3.98 -- 135.6  &  51.8 & 1.870\,$\times$\,1.300 & $-16.1$ & 2.57 & this work  \\
 90.700 & OVRO\footnote{L, E, H, and UH configurations.}+NMA\footnote{D,
     C, and AB configurations.} & 2.90 -- 139.0 & 69.7 & 1.52\,$\times$\,1.51  & $-9.6$ & 2.72 & paper I \\
335.416 & SMA\footnote{Extended and Compact configurations.} & 13.3 -- 241.5 & 15.5 & 0.85\,$\times$\,0.78  &  $-32$  & 34.9 & this work \\
\hline
\end{tabular}
\end{minipage}
\end{table*}

With all this in mind, we performed a detailed study of the \gn\ high-mass
star forming region (SFR) which is known to harbor several UC \hii\ regions
and a HMC (Garay et al. 1985, 1998; \cite{wc89}; \cite{fc00}; \cite{rsf05a}, hereafter
paper I).  The HMC was firstly identified by Garay et al. (1998) in the \amm
(2,2) inversion transition.  A summary of molecular line observations towards
the HMC up to 2004 is given in Remijan et al. (2004; see references therein).  
No unambiguous proof of the presence of OB-type (proto)stars inside the HMC, 
such as the detection of free-free emission, has been found yet.
However, the detection of 
\wat\ (\cite{hc96}, hereafter HC96; \cite{fc00}) 
and OH (\cite{garay85}) maser emission is
considered evidence of their existence. 
All these features make the
\gn\ star forming region an ideal target to study the early phase of
the high-mass star formation process.\par

Based on lower quality images of the region, in paper~I, we estimated the
lifetime ratio between the UC \hii\ and the HMC phases, concluding that
the former should last $\sim$3 times longer than the latter.
With the new observations presented here, we wish to set tighter
constraints on the statistical study of paper~I and improve our knowledge
of the HMC. In this paper we will focus on the latter issue, while a forthcoming
paper will be devoted to a more general analysis of the high-mass young
stellar cluster in the region.\par

It is worth noting that the \gn\ SFR was believed to be located at the near
kinematical distance, based on H$\alpha$ emission and H$_2$CO absorption line
observations towards the UC \HII\ regions.  This is why in the literature one
finds distance estimates of 3.8 kpc (\cite{gg76}), 4.5$\pm$1 kpc
(\cite{downes80}), and 3.5 kpc (Churchwell, Walmsley \& Cesaroni 1990).
However, interferometric observations of the HI line at 21\,cm seen in
absorption against the \HII\ regions (Kolpak et al. 2003) have established
that the region is located at the far kinematical distance of
12.6$\pm$0.3~kpc.  A similar result has been obtained by other authors
(Pandian, Momjian \& Goldsmith 2008), who derive a value of 11.8$\pm$0.5\,kpc
or 12.0$\pm$0.4\,kpc, depending on the adopted rotation curves.  Note that
the latter measurement was done towards the center of the nearby
G\,19.61-0.13 region, which is offset by 11\arcmin\ from the HMC position.
For this reason, in this paper, we prefer to adopt $d =$\,12.6~kpc which
comes from a measurement made towards the center of the region of interest
for us (\gn).


Finally, we point out that recently, Wu et al. (2009) have
reported on the detection of inverse P-Cygni profiles in \tCO\ $J=3-2$ and CN
$J=3-2$ lines towards the \gn\ HMC.  This result was obtained using the
Submillimeter Array (SMA) archive data originally taken by us.  Wu et al.
(2009) conclude that such profiles are due to infall motions inside the HMC.
In this paper, we improve on this result presenting a larger study of the
line emission from the HMC.

\section{Observations and Data Reduction}
\label{s:o}

Aperture synthesis observations of the continuum and molecular line
emission towards the G19.61--0.23 star forming region,
from centimeter (cm) to sub-millimeter (submm) wavelengths,
were carried out using the
Very Large Array (VLA) of the
National Radio Astronomy Observatory 
(NRAO)\footnote{The National Radio Astronomy Observatory is 
a facility of the National Science Foundation operated under 
cooperative agreement by Associated Universities, Inc.}
(Sect.~\ref{sss:obs_vla}), the
Nobeyama Millimeter Array (NMA) of the 
Nobeyama Radio Observatory\footnote{The Nobeyama Radio Observatory 
is a branch of the National Astronomical Observatory of Japan, 
operated by the Ministry of Education, Culture, Sports, Science and Technology, Japan.}
(Sect.~\ref{sss:obs_nma}), the
Owens Valley Radio Observatory (OVRO)\footnote{Research at 
the Owens Valley Radio Observatory is supported by the National Science Foundation 
through NSF grant AST 02-28955.} millimeter array
(Sect.~\ref{sss:obs_ovro}),
and
the Submillimeter Array (SMA)\footnote{The Submillimeter Array 
is a joint project between the Smithsonian Astrophysical Observatory 
and the Academia Sinica Institute of Astronomy and Astrophysics 
and is funded by the Smithsonian Institution and the Academia Sinica.}
(Sect.~\ref{sss:obs_sma}),
in the period from 2002 to 2007.
We summarize our continuum imaging and spectral line observations
in Tables \ref{tbl:contobs} and \ref{tbl:lineobs}, respectively.

\begin{table*}
\begin{minipage}[t]{\textwidth} 
\caption{Summary of Interferometric Molecular Line Observations}
\label{tbl:lineobs}
\renewcommand{\footnoterule}{}  
\begin{tabular}{lcccccc}
\hline \hline
\lw{Line} & \lw{$f_{\rm rest}$} & \lw{Array} & \multicolumn{2}{c}{Synthesized Beam} & \lw{$\Delta v$\footnote{Effective velocity resolution.}} & Image\footnote{Typical RMS noise level per velocity channel.} \\
\cline{4-5}
          &                     &            & $\theta_{\rm major}\times\theta_{\rm minor}$ & P.A. &                  & Noise Level \\
          & (MHz)               &            & (arcsec)   & (deg)                                  & (\kms )          & (mJy beam$^{-1}$) \\
\hline
\wat\ maser & 22235.080 & VLA-B &  0.32\,$\times$\,0.23    & $-9.6$ &  0.66 & 90 \\
\HtCOp\ (1--0)        &  86754.328  & OVRO  & 4.00\,$\times$\,3.13    & $-21$ &  0.43 & 61   \\
SiO (2--1) $v=0$      &  86846.998  & OVRO+NMA  & 3.67\,$\times$\,2.40    & $-3.0$ &  0.90 & 38 \\
\tCO\ (3--2)          & 330587.9601 & SMA   & 0.94\,$\times$\,0.83        & $-27$ &  0.40 & 340   \\
\CeO\ (3--2)          & 329330.5453 & SMA   & 0.94\,$\times$\,0.85        & $-22$ &  0.80 & 338 \\
\mcn\footnote{\mcniso\ $18_K-17_K$ lines as well, see Fig.~ \ref{fig:SPmcn}.} $18_K-17_K$ & 331071.594\footnote{For \mcn\ $K=$\,0}  & SMA   & 1.45\,$\times$\,1.38 & $-16$ &  1.0 &   145 \\
\hline
\end{tabular}
\end{minipage}
\end{table*}
%

\subsection{VLA Observations} 
\label{sss:obs_vla}
We performed VLA observations of the continuum emission at 
6\,cm, 1.3\,cm, and 7\,mm as summarized in Table \ref{tbl:contobs}.
For all the observations described below, 
we used quasars J1832$-$105 as a phase-calibrator and
J1331+305 and/or J0137+331 as flux- and bandpass-calibrator(s).

\paragraph{6\,cm --- }
The A- and C-array observations were done on November 1, 2004 (project code AF\,415)
and July 11, 2005 (AF\,422), respectively, with the standard correlator
configuration for continuum imaging
providing a 172\,MHz bandwidth with dual polarization.

\paragraph{1.3\,cm --- }
We performed the B-array observations on May 28 and 30, 2005 (AF\,415),
and the C-array observation on November 18, 2006 (AF\,422).
We observed the continuum emission in the ``BD''
intermediate-frequency (IF) pair, with 25\,MHz 
bandwidth, and the \wat\ maser emission in the ``AC'' pair 
with 3.125\,MHz bandwidth and 64-channels,
providing a velocity resolution of 0.66 \kms.
Because of hardware limitations during the EVLA transition phase, 
we manually supplied the sky-frequency of the maser line at the beginning of
each switching cycle (12 minutes) instead of using Doppler tracking.
We excluded all the LHCP data taken at the B-array observations due
to a 180\deg\ phase jump over the correlator band.
Furthermore, all the data taken with five EVLA antennas during the
D array observations were excluded due to error on the amplitude calibrations.
We have produced continuum images by merging the B- and D-array visibility data
(Table \ref{tbl:contobs}).
For the water maser emission, the two data sets were not merged due to
variability between the B- and D-array observations. We thus prepared a
a 3D data cube of the maser line emission using only the B-array data,
which allowed us to attain the best
angular resolution.\par

\paragraph{7\,mm --- }
The 7\,mm observations were performed on May 8, 2007 with the D array,
employing the fast-switching technique. The adopted switching cycle 
consisted of a 2.0-minute integration on the target and a 50-second
integration on the calibrator.
Before making an image, 
we excluded all the data taken with the nine EVLA antennas due to an unknown error 
on the amplitude calibrations, leaving us with only 17 usable antennas.

\subsection{NMA Observations}
\label{sss:obs_nma}

The NMA observations towards \gn\ were carried out in the period from
December 2002 to May 2003 with 3 array configurations (D, C, and AB).  
We observed the SiO (2--1) $v=0$ line in the upper-side band (USB) 
using the FX correlator
with 32\,MHz bandwidth centered at the line frequency,
giving a velocity resolution of 0.108~\kms.
For the continuum and line emission, 
we used the Ultra Wide Band Correlator (UWBC)
with a narrow 512\,MHz bandwidth in each sideband.
Although this correlator configuration should allow us to observe
a total bandwidth of 1\,GHz with the dual side bands, 
after removing the channels affected by line
emission, the effective bandwidth usable for continuum emission measurements
was $\sim$450\,MHz.
We used 3C\,273 as passband calibrator and J1743$-$038 as phase and gain calibrator.  
The flux densities of J1743$-$038 were bootstrapped from Uranus, 
and the uncertainty is estimated to be 10\%.  
All the data were calibrated and edited using the UVPROC2 and MIRIAD packages.

\subsection{OVRO Observations}
\label{sss:obs_ovro}

The OVRO observations of \gn\ were carried out in the period from 
September 2003 to May 2004 in the three array configurations E, H, and UH.
We observed
the \HtCOp\ (1--0) and SiO (2--1) $v=0$ lines in the lower side-band (LSB).  
For the continuum emission, we simultaneously used the Continuum 
Correlator with an effective bandwidth of 4\,GHz and
the newly installed COBRA correlator with 8 GHz bandwidth.  
The line contamination could not be estimated
as these correlators do not have spectroscopic capabilities.
For molecular lines, we used the digital correlator with 
15\,MHz bandwidth and 60 channels for the SiO line,
and 7.75\,MHz and 62 channels for the \HtCOp\ line.
We used 3C\,273 and 3C\,454.3 as passband calibrators, 
and J1743$-$038 as phase and gain calibrators.
The flux densities of NRAO\,530 and J1743$-$038 were determined from 
observations of Uranus and Neptune.  
We estimate the uncertainty of the flux calibration to be 10\%.  
The data were calibrated and edited using the MMA and MIRIAD packages.\par

\subsection{Combining NMA and OVRO Data: The SiO (2--1) $v=0$ Line}
\label{sss:obs_combine}
We combined the interferometric SiO (2--1) $v=0$ data taken with the
NMA (Sect.~\ref{sss:obs_nma}) and OVRO (Sect.~\ref{sss:obs_ovro}) interferometers;
the diameters of the element antennas are 10.0\,m and 10.4\,m, respectively.
For this purpose, we adopted the observational
set-ups adopted for both data sets were as similar as possible,
with two important differences, though:
the frequency resolution and the integration time for a single visibility.
We therefore smoothed the NMA visibilities every 8 channels, 
and averaged them so to achieve a 3.35-seconds integration time per visibility.
We verified that the visibilities taken with the two arrays
have a fairly nice consistency in flux calibration when 
we compared them over the common range of the projected
baseline length.
Subsequently, we produced a 3D cube using
task IMAGR of the AIPS package with a
robust parameter of +1 to find a compromise between angular resolution
and image-fidelity.

   \begin{figure*}
   \centering
   \includegraphics[angle=0,width=15cm]{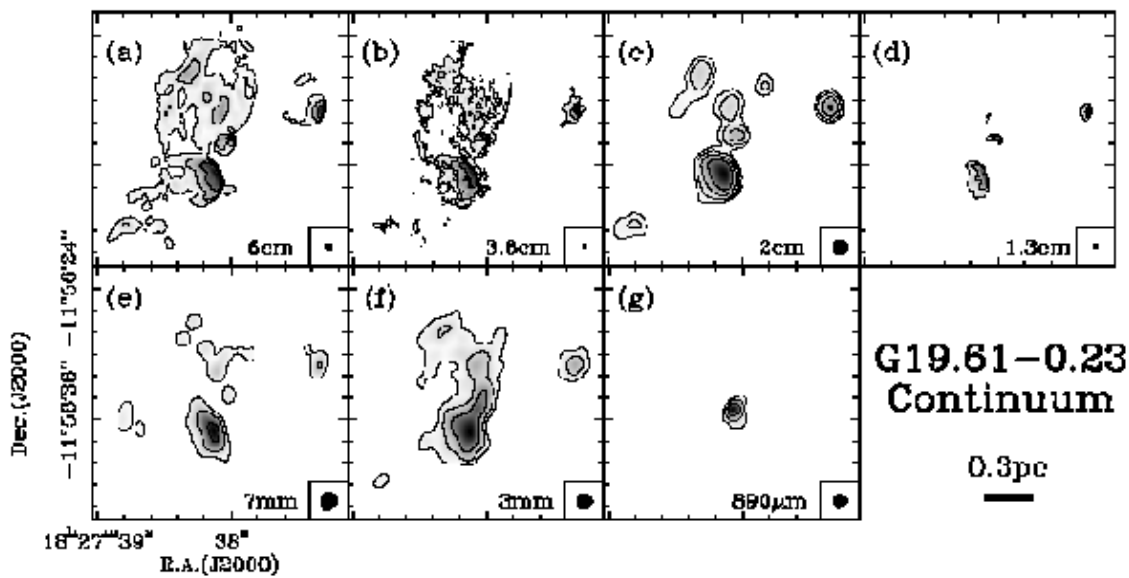} 
      \caption{Continuum emission maps toward the \gn\ high-mass star forming region.
All the contour levels are drawn with $\pm 7\sigma\cdot 2^n$ where
$\sigma$ is the RMS noise level of the image (Table \ref{tbl:contobs}) 
and $n=0,1,2,\dots$.
The wavelength and FWHM of the synthesized beam are indicated
in the bottom right corner of each panel.
See Table \ref{tbl:contobs} for the values of the synthesized beams.
              }
         \label{fig:contmaps}
   \end{figure*}

\subsection{SMA Observations and Data Reduction}
\label{sss:obs_sma}

Aperture synthesis continuum and line emission observations of
\gn\ at 890\,\um\ were
carried out with the SMA on May 12, 2005 in the compact array 
configuration and on July 8, 2005 in the extended configuration.
The shortest projected baseline length, i.e. the shadowing limit, 
was about 11.9\,m.  
This makes our SMA observations insensitive to structures more 
extended than 15\farcs5, corresponding to 0.95\,pc at a distance of 12.6\,kpc.
The SIS receivers were tuned at a frequency of 335.4158\,GHz
to observe the \tCO\ (3--2), \CeO\ (3--2), and \mcn\ (18--17) lines
in the LSB, and the
HC$^{18}$O$^+$ (4--3) and CN (3--2) lines in the USB.
Each side-band covers a 2\,GHz bandwidth.
The attained synthesized beam sizes,
effective velocity resolution after smoothing, 
and image sensitivity are summarized
in Tables \ref{tbl:contobs} and \ref{tbl:lineobs}.\par

We used 3C\,454.3 as bandpass calibrator, 
and J1743--038 and J1924--292
as phase and gain calibrators.
Flux densities of the two calibrators were bootstrapped
from observations of Uranus and Neptune
(J1743--038: 1.64 Jy in May and 1.55 Jy in July;
J1924--292: 3.16 Jy in May and 3.60 Jy in July), and
were stable within 12\% during the observing period.
We estimate a final uncertainty on the flux calibration of $\sim 20\%$.
The data calibration was done using the MIR and MIRIAD packages.
Following standard calibration procedures,
we tentatively subtracted the continuum emission from the visibility data, 
then imaged the continuum in each side-band.
After verifying that the continuum emission is confined within 
a compact region at ($\Delta\alpha$, $\Delta\delta$)
$= (-1\farcs2, +2\farcs2)$ with respect to the phase tracking center
(PTC; ~R.A. = 18$^h$ 27$^m$ 38.15$^s$, 
Dec. = $-11$\deg 56\arcmin 39\farcs50 in J2000),
we subtracted the continuum contribution from the
visibility data by giving the position offset 
in task UVLIN in MIRIAD package.
For the final continuum subtraction process, 
we identified, at least, 
33 and 24 lines in the USB and LSB, respectively.
Subsequently, we constructed continuum emission images 
with natural and uniform visibility weighting functions
in conjunction with the multi-frequency synthesis technique.
For the continuum imaging, 
we did not apply another visibility weighting based 
on the system temperature (\Tsys ) of each element antenna,
because the \Tsys\ information in the USB for 
the compact array observations were not recorded properly.
In contrast, the \Tsys-based visibility weighting
was applied to the line data in the LSB, for which the \Tsys\ was properly
recorded.
Note that the \Tsys\ problem affecting the USB was not
realized by Wu et al. (2009), who thus produced the incorrectly weighted,
low-resolution continuum image shown in their Fig.~1.

Atmospheric seeing was estimated
using the continuum emission maps of the (point-like) calibrators,
because the apparent angular diameter of the calibrators is related
to the size scale characterizing 
the atmospheric turbulence.
The time-averaged size scale over each observing track should be
comparable to the beam-deconvolved diameters of the calibrators.
By this means, we estimate the average seeing for J1743--038
to be 0\farcs15 and 0\farcs95 for the extended and
compact configuration observations, respectively.
We found that the seeing towards J1924--292 was twice
worse than that towards J1743--038 in the extended array observations.
In addition, given the fact that the angular distance between J1924--292
and the PTC is rather large (21.8\deg)
we decided not to use this quasar as phase calibrator.
Finally, we reconstructed all the images of the 
continuum and line emission data.\par

\section{Results and Analysis}
\label{s:results}

\subsection{Continuum Emission}
\label{ss:r_cont}
\subsubsection{Overview of the \gn\ Star Forming Region}
\label{sss:contoverall}

Fig.~\ref{fig:contmaps} presents the continuum emission maps,
including already published 3.8\,cm and 3\,mm images.
As discussed in paper~I, the cm maps, representing free-free emission,
show that the region contains a cluster of UC \hii\ regions, i.e., 
young massive stars.
One can see that the overall morphology of the cm emission does not change
much with wavelength.
Most of the apparent differences are due to different angular resolution
and sensitivity to extended structures
(see Table $\ref{tbl:contobs}$).
The 7\,mm and 3\,mm maps, which have comparable angular resolutions, 
also show similar structures.
Such a similarity implies that these two images contain
both thermal dust emission and optically thin free-free emission, as
discussed for the 3\,mm map in paper~I.
Note that the compact 7\,mm emission to the east of UC \hii\ region A
(see paper I for labeling of the UC \hii\ regions) is very likely
an artifact due to inadequate sampling of the visibility plane
(Sect.~\ref{sss:obs_vla}).
The SMA 890~\um\ image looks significantly different 
from the cm- and mm maps, as it shows a bright compact source towards the HMC position 
previously reported.
No other 890 \um\ sources than the HMC are detected 
over the SMA FoV, above a 3\sgm\ upper limit of 105 mJy \pbeam,
corresponding to a brightness temperature in the synthesized beam of
\Tsb(3\sgm ) = 0.22\,K.
In the following, we describe the continuum results obtained at
890 \um\ (Sect.~\ref{sss:smacont}) and the
high-resolution images at cm wavelengths (Sect.~\ref{sss:vlacont}).

   \begin{figure*}
   \centering
   \includegraphics[angle=0,width=13cm]{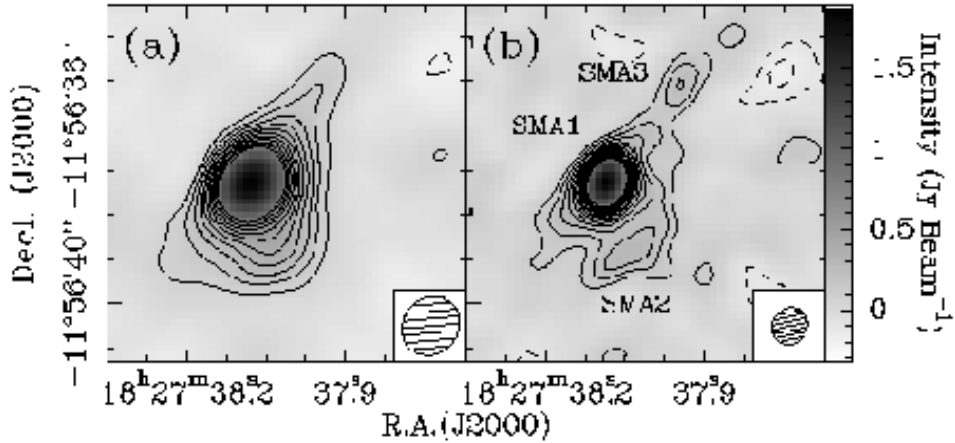} 
      \caption{Continuum emission maps at 890 \um\ taken with SMA toward \gn\ with
(a) natural weighting and (b) uniform weighting.
The solid contours start at the $3\sigma$ level and increase in steps of $+2\sigma$ 
up to the 50\% level for a clarity of the maps.
The dashed contours start at the $-3\sigma$ level and decrease in steps of $-2\sigma$.
The image noise levels are 54.1 and 34.9 mJy beam$^{-1}$ for panels 
(a) and (b), respectively. The sensitivity of the uniformly weighted map
is better than that of the naturally weighted map.
We believe that this is due to the fact that visibilities have not been
weighted taking into account the
system temperature of each antenna (see Sect.\ref{sss:obs_sma}).
The synthesized beams are shown in the bottom right corners
(1\farcs38$\times$1\farcs22 with P.A.$=-44$\deg\ for natural weighting;
0\farcs85$\times$0\farcs78 with P.A.$=-32$\deg\ for uniform weighting).
              }
         \label{fig:smacont}
   \end{figure*}

\begin{table*}
\begin{minipage}[ht]{\textwidth}
\caption{Results of Submm Continuum Emission Observations}
\label{tbl:smacont}
\renewcommand{\footnoterule}{}  
\begin{tabular}{ccccccccc}
\hline\hline
\lw{Name} & \multicolumn{2}{c}{Peak Position (J2000)} & 
$I_{890\mu m}$\footnote{Peak intensity. The uncertainty may be given by the 3\sgm\ 
level of the image noise level, i.e., 0.1 Jy \pbeam .} & 
$S_{890\mu m}$\footnote{Flux density, see Sect.~\ref{sss:smacont}.} & 
$D_{\rm d}$\footnote{Effective diameter of the submm emission, 
which is given by $2\sqrt{A/\pi}$ where $A$ is the area enclosed by the 
5\sgm\ level contour enclosing each source.
Notice that, at SMA2 and SMA3, 
the 5\sgm\ level corresponds to the 61\% and 71\% levels 
with respect to their peak intensities, respectively.} & 
$R_{\rm d}$\footnote{Effective radius of the submm emission.} &
$M_{\rm d}$\footnote{Mass of the core estimated from the flux
$S_{890\mu m}$ in Col.~5, assuming $\beta = 1$ and a dust temperature 
of 80\,K.} \\
\cline{2-3}
          & R.A.(hh:mm:ss) & Dec.(dd:mm:ss) & (Jy \pbeam ) & (Jy) & (arcsec) & (pc) & (\Msun ) \\
\hline
SMA1  & 18:27:38.069 & $-$11:56:37.30 & 1.89 & 3.2  & 2.4 & 0.072 & 1300 \\
SMA2  & 18:27:38.021 & $-$11:56:38.75 & 0.29 & 0.52 & 1.5 & 0.045 & 220 \\
SMA3  & 18:27:37.955 & $-$11:56:35.08 & 0.25 & 0.15 & 0.81 & 0.025 & 60 \\
\hline
\end{tabular}
\end{minipage}
\end{table*}

   \begin{figure*}
   \centering
   \includegraphics[width=10.0cm,angle=0]{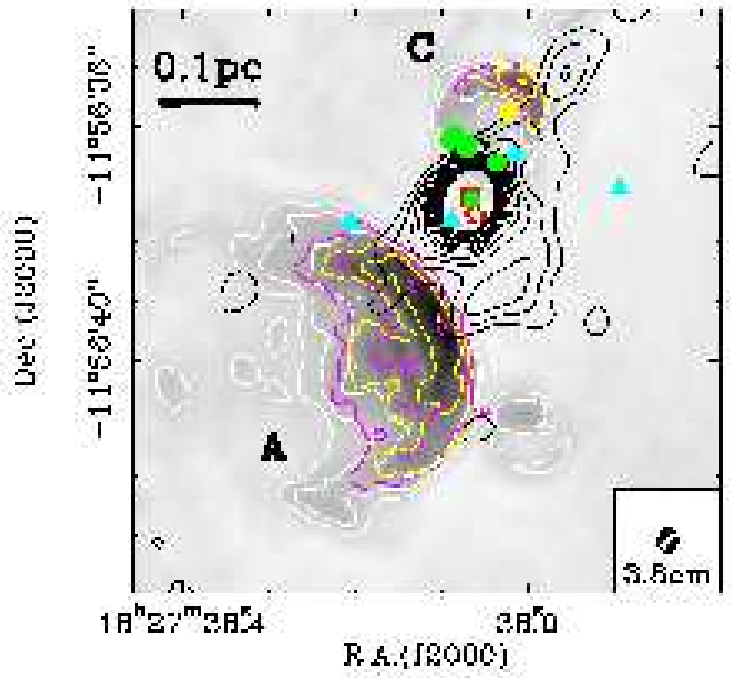} 
      \caption{Comparison of the 6\,cm (greyscale plus white contour),
3.8\,cm (magenta contour), and 1.3\,cm (yellow contour) continuum
emission maps generated with the common minimum UV distance of 13.3
$k\lambda$ (see Sect.~\ref{sss:vlacont}).
Contour intervals for the cm maps and the submm map 
are the same as those in Figs.~\ref{fig:contmaps} and \ref{fig:smacont}, respectively,
where the 1\sgm\ noise levels are 
0.23, 0.18, and 0.36 mJy \pbeam\ and
the beams 0\farcs433$\times$0\farcs321,
0\farcs262$\times$0\farcs195, and
0\farcs344$\times$0\farcs251,
for the 6.0, 3.8, and 1.3\,cm maps, respectively.
The largest of the three beams is shown in
the bottom right.
The filled red rectangular and the double red circles indicate
the peak positions of the 3\,mm (paper I) and
890\,\um\ (Sect.~\ref{sss:smacont}) continuum emission, respectively.
The filled green circles and light-blue triangles show the
positions of the OH (\cite{garay85}) and \wat\ (\cite{hc96}) masers, respectively.
Labels A and C refer to the UC \hii\ regions 
(notation as in paper I and references therein).
The filled yellow circle associated with the UC \hii\ region C
indicates the peak position of the isolated 7\,mm emission 
seen in Fig.~\ref{fig:contmaps}e close to the HMC.
              }
         \label{fig:cmmaps}
   \end{figure*}

\subsubsection{890 \um\ Continuum Emission}
\label{sss:smacont}
In paper I, we have demonstrated that the 3\,mm continuum emission towards
the HMC is mostly due to thermal dust emission with a small contribution from
free-free emission.  The latter is instead insignificant in the newly obtained
SMA image at 890 \um, which is dominated by the dust continuum emission.
Fig.~\ref{fig:smacont} shows a close-up view of the SMA 890 \um\ images
produced with natural and uniform visibility weighting functions.  It is
noteworthy that the continuum emission in the natural weighted map
(Fig.~\ref{fig:smacont}a) is elongated to both the south and northwest.  The
uniformly weighted map (Fig.~\ref{fig:smacont}b) clearly resolves the emission
and shows that the elongation is attributed to two additional weak sources,
whose peak intensities are slightly above the 7\sgm\ level.  These results
indicate that there are (at least) three submm sources in the region.
Hereafter, we will name these SMA1 (the HMC), SMA2 (the core to the south), and
SMA3 (the core to the north-west), as indicated in Fig.~\ref{fig:smacont}b.

The most intense 890 \um\ source, SMA1, is located 0\farcs32 northeast of the
peak of the 3\,mm ``dust'' continuum emission 
(see Fig.~2c in paper I and Fig.~\ref{fig:cmmaps} of this work).
This means that the 890\,\um\ and 3\,mm continuum peaks coincide within the
errors. Therefore, in this study we will identify SMA1 with the HMC, whose
position is obtained from the 890 \um\ image (Table \ref{tbl:smacont}).  The
peak intensity ($I_{\nu}$) of the 890 \um\ continuum emission is 1.89 Jy
\pbeam, corresponding to \Tsb\ = 31.0\,K.
The spectral index ($\alpha$) between 3\,mm and 890 \um\ is 
$\alpha_{\rm 3mm-890 \mu m}\gtrsim 2.7$, 
implying a power-law exponent for the dust emissivity, 
$\beta$, of $\gtrsim 0.7$, assuming optically thin dust emission
and the Rayleigh-Jeans approximation.  
For this estimate, 
we integrated the 890\,$\mu$m flux over the three sources SMA1, SMA2, and
SMA3 and used the
continuum flux at 3\,mm (147~mJy) obtained from the data of
paper I, where the free-free continuum contribution from the nearby UC~\hii\
region was subtracted by extrapolating the 1.3\,cm continuum map to 3\,mm
under the assumption of optically thin emission.
We stress that the latter flux is affected by significant
uncertainties due to the method adopted to subtract the free-free contribution
(see paper~I).
Therefore, in the following
we prefer to adopt $\beta=1$, which is consistent with the value derived
above, within the uncertainty, and falls in the
range $\beta=1$--2 found in the literature.\par

The 890 \um\ flux densities (\Snu ) in Table~\ref{tbl:smacont} 
allow us to estimate the gas-plus-dust mass
traced by thermal emission (\Mdust). 
For this purpose we have assumed a dust temperature (\Tdust ) of 80\,K
(paper I)  and have calculated the values of \Snu\ by
integrating the emission in Fig.~\ref{fig:smacont}b
over the regions inside the 5\sgm\ contour levels of the sources.
The value \Tdust $\simeq 80$\,K was obtained from the fit to the continuum
spectrum
(see paper I), from the cm to the mid-infrared regime,
including the newly obtained 7\,mm and 890 \um\ fluxes.
We estimate that the ambiguity in defining the boundary between SMA1 and
SMA2 causes a $\sim 5\%$ error on the value of \Snu.
The resulting masses are given in Table~\ref{tbl:smacont}.
Note that these values are approximately inversely proportional to the
dust temperature and are thus affected by the uncertainty on this
latter parameter accordingly. 
Since the assumed temperature is likely to
be correct within a factor 2, 
we believe that the estimated masses are
also affected by a similar uncertainty.\par

%
   \begin{figure*}
   \centering
   \includegraphics[width=10.0cm,angle=0]{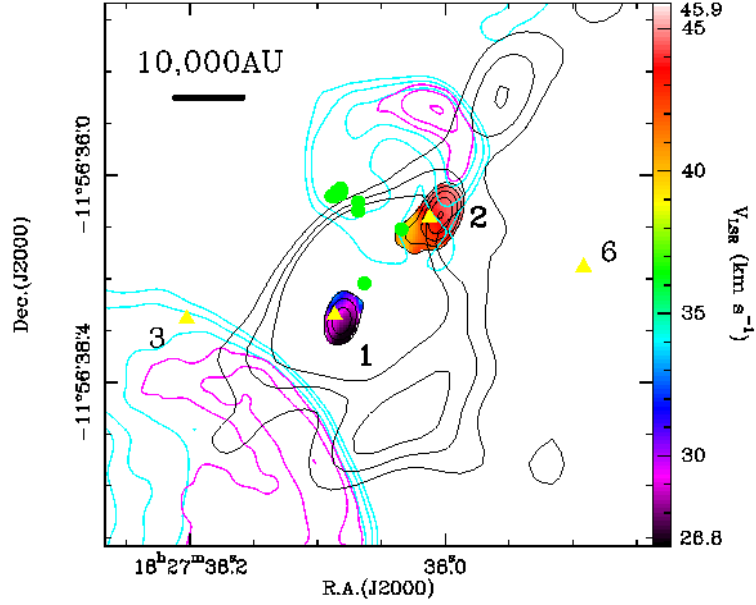} 
      \caption{
First moment map of the \wat\ masers (color image) and maps of the
continuum emission (color contours). The black contours overlayed on
the maser velocity map correspond to the
integrated intensity of the maser spots,
with levels starting from the 5\sgm\ level with 5\sgm\ steps.
The black, cyan, and magenta contours represent the first 3 contours of
the 890 \um, 6\,cm, and 3.8\,cm continuum emission, respectively,
(the same as in Fig.~\ref{fig:smacont}).
The yellow triangles and green-filled circles
show, respectively, the positions of the \wat\ (\cite{hc96}) and OH (\cite{garay85}) masers.
The numbers associated with the yellow triangles are used to identify
the maser features as done by Hofner \& Churchwell (1996).
              }
         \label{fig:masermap}
   \end{figure*}

%
   \begin{figure}[b!]
   \centering
   \includegraphics[width=6.0cm,angle=0]{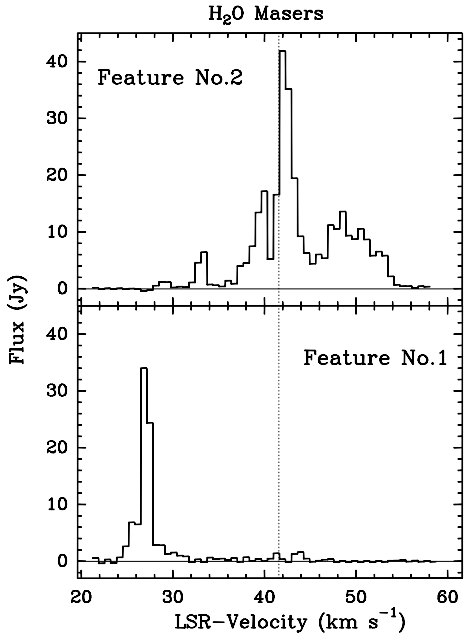} 
   \caption{\wat\ maser spectra toward the maser features No.1 and 2
obtained by integrating the emission inside the 5\sgm\ contour of
each emission (see Fig.~\ref{fig:masermap}).
The vertical dashed line indicates the systemic velocity (\Vsys ) of the
cloud, \Vlsr $=$\,41.6\,\kms.
        }
         \label{fig:masersp}
   \end{figure}

\subsubsection{Centimeter Continuum Emission towards the HMC}
\label{sss:vlacont}

To assess the presence of cm emission towards the HMC,
we reconstructed the VLA 6\,cm, 3.8\,cm, and 1.3\,cm continuum 
images in Fig.~\ref{fig:cmmaps} by making use of visibilities
whose minimum spatial frequency range is set to be the same for the 3 bands.
We chose the shortest baseline equal to 13.3\,k$\lambda$, 
namely that of the 890 \um\ maps (see Table \ref{tbl:contobs}). This allows one
to resolve out the extended emission.
Fig.~\ref{fig:cmmaps} clearly shows that 
no free-free emission is detected towards the peak positions of the 
three 890\,\um\ sources. Note that SMA1 is located 
at the center of the \wat\ and OH masers' distributions, whereas
no \wat\ and OH maser spot appears to be associated with SMA2 and SMA3.
The upper limits (3\sgm) obtained from the cm images correspond to 
brightness temperatures
over the synthesized beam (\Tsb ) of 199, 230, and 30.9\,K,
respectively at 4.86, 8.42, and 22.27\,GHz
(see the caption of Fig.~\ref{fig:cmmaps} for the corresponding noise
levels and beam sizes).\par

We point out that the 7\,mm emission seen to the north of the HMC
(see Fig.~\ref{fig:contmaps}e)
coincides with the cometary UC \hii\ region C (Fig.~\ref{fig:cmmaps}).
We thus argue that free-free emission from this UC \hii\ region would
significantly contribute to the corresponding 7\,mm continuum emission,
whose peak intensity is 25.3 mJy \pbeam\ ($\sim$10 \sgm).
It is also interesting to note that SMA3 lies right in front of the vertex
of the cometary shaped UC \HII\ region C, suggesting a physical connection
between the two. This could explain the cometary shape, with the existence of
dense material preventing expansion of the ionized gas towards north-west.

\subsection{\wat\ masers}
\label{ss:maser}

\subsubsection{Overall Results}
\label{sss:maser_results}

Fig.~\ref{fig:masermap} shows the \wat\ maser distribution with respect to
the UC \HII\ regions.
The color image is a map of the first-order moment of the maser lines.
We detected two out of the six ``maser features'' previously reported by HC96, 
i.e. features N.\,1 and 2 in their notation.
Here, we use the term ``maser feature'' for a well-defined,
spatially isolated group of maser spots
(see, e.g., HC96 and \cite{rsf05b}).
With an angular resolution of 0\farcs3, one cannot distinguish the maser
spots associated with all the different lines seen in the spectrum
(Fig.~\ref{fig:masersp}).
The fact that no maser emission was detected towards the other four features
identified by HC96, must be due to the high-variability of H$_2$O masers,
because the intensities measured by HC96 are all well above our
3\sgm\ sensitivity of 135~mJy \pbeam.

Figs.~\ref{fig:masermap} and \ref{fig:masersp} clearly show that
feature 1 presents only blueshifted emission,
whereas feature 2 is mostly redshifted, 
despite the presence of a few
blueshifted lines.
Notwithstanding the well known high variability of water masers, we note that
these two features have persisted with approximately the same spectral shape
since December 1991, when the VLA observations of HC96 were made.
The terminal velocities (\Vt) at which blue- and red-shifted maser emissions
are seen towards features 1 and 2 differ by $\sim$15--20 \kms\
with respect to the systemic velocity (\Vsys) of 41.6 \kms. 
All this suggests that the maser emission may be originating
in a bipolar jet driven by a putative young stellar object (YSO) inside the HMC.
We will come back to this issue in Sect.~\ref{sss:HtCOp}.

\subsubsection{Origin of the \wat\ Maser Features 1 and 2}
\label{sss:maserSMA1}

Features 1 and 2 (see Figs.~ \ref{fig:masermap} and \ref{fig:masersp}) seem
to have persisted more than a decade.  From their distribution (see Fig.~3 of
HC96), one can argue that only these two features out of the six reported by
HC96 are associated with SMA1.

Interferometric observations of \wat\ masers at 22\,GHz indicate that
they are likely excited in shocked regions at the interface between
(proto)stellar jets and the ambient gas (e.g., \cite{rsf00}),
although in some objects these masers have been suggested to be tracing
rotating disks (e.g., \cite{torrelles96}).
If the latter were the case of the masers in SMA1, from the separation
between the blue- and red-shifted features
(1\farcs70) and their relative velocities ($\sim$35\,\kms)
one could estimate the mass needed to
ensure centrifugal equilibrium: this is $\sim 3700$ \Msun, 
which is $\sim$3~times
greater than that obtained from the submm continuum emission
(see Table \ref{tbl:smacont}). This suggests that rotation is not a viable
explanation for the kinematics of the masers in the HMC. Moreover,
as we will see in Sect.~\ref{sss:HtCOp}, the \HtCOp(1--0) line appears
to trace a bipolar outflow oriented SE--NW.
In this scenario the two H$_2$O maser features can be associated with a
bipolar jet feeding the outflow. We thus conclude
that the jet interpretation is more likely.

   \begin{figure*}[ht!]
   \centering
   \includegraphics[angle=0,width=11cm]{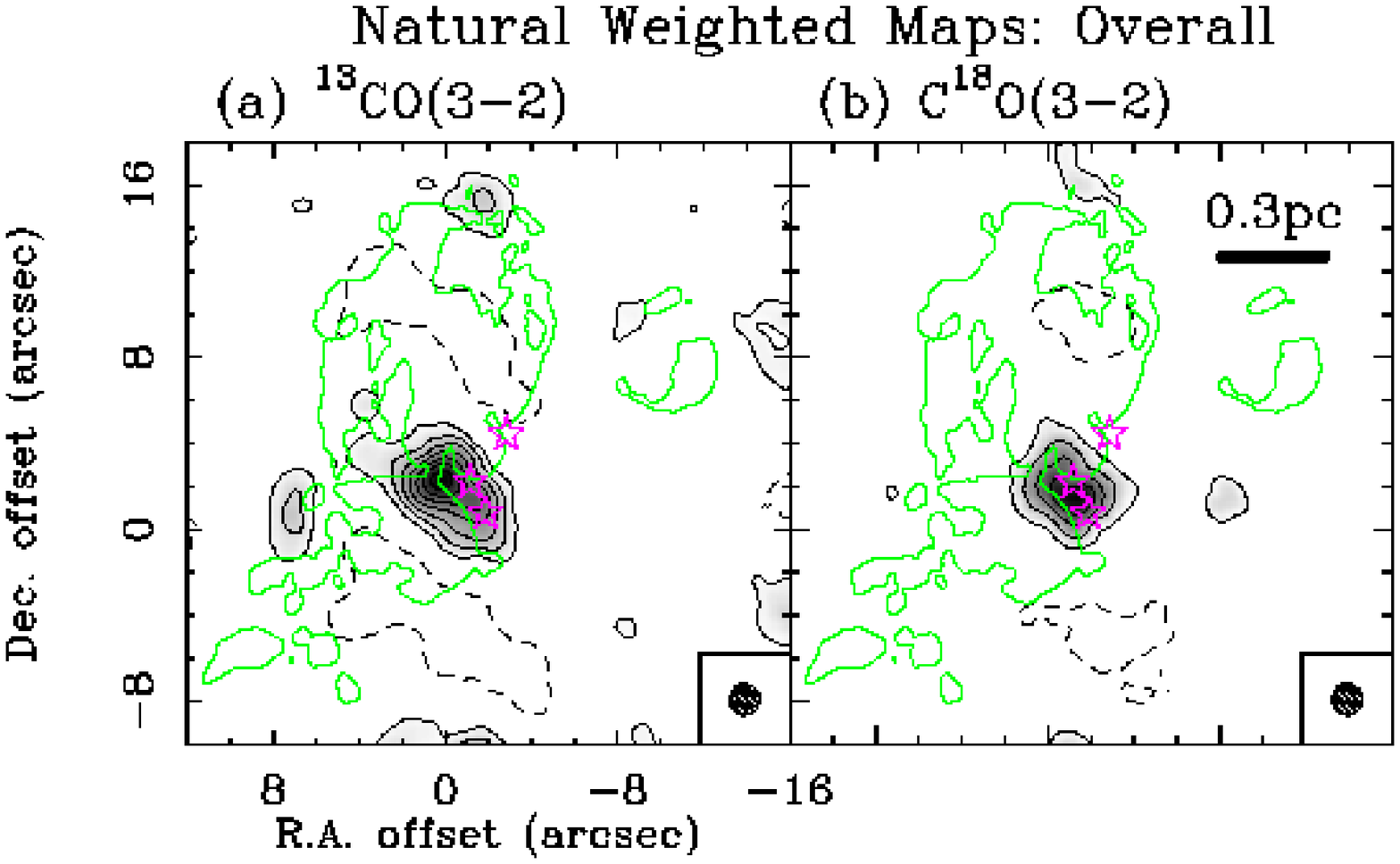} \\ 
   \includegraphics[angle=0,width=11cm]{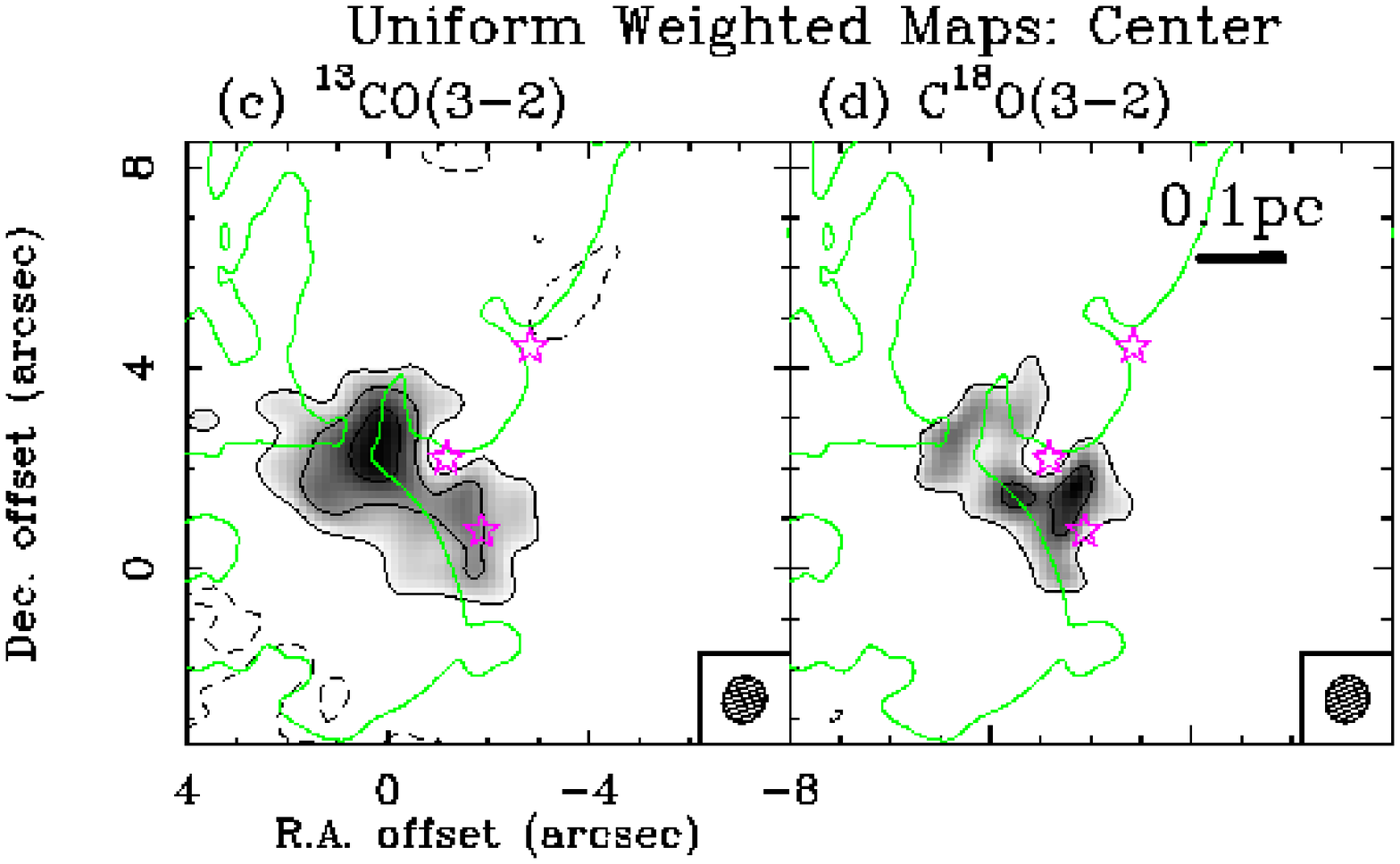} 
      \caption{Total integrated intensity maps of the 
\tCO\ (3--2) ({\it left panels}) and \CeO\ (3--2) ({\it right panels}) 
lines produced with natural ({\it upper panels}) and uniform ({\it lower panels}) visibility
weighting functions.
Note that the upper panels show the whole area of the \gn\ star forming region,
while the lower panels magnify the region centered on the HMC
(greyscale plus thin contour; see Table \ref{tbl:lineobs}).
For a comparison purpose, the 7\sgm\ level contour of the
6\,cm continuum (i.e., free-free) emission as in
Fig.~\ref{fig:contmaps}a is shown by thin green contours.
The magenta stars in the lower-panels indicate the peak positions
of the three 890 \um\ continuum sources (Table \ref{tbl:smacont}).
The contours of the CO isotopomer maps start
from the 5\sgm\ level in steps of
3\sgm, where the 1\sgm\ noise levels are
1.9 and 1.8 Jy \pbeam\ \kms\ for the natural and uniform weighted
\tCO\ maps, and 
0.8 Jy and 1.7 \pbeam\ \kms\ for the natural and uniform weighted
\CeO\ maps, respectively.
We have integrated the emission over the LSR-velocity ranges
30.2 $\leq$ \Vlsr /\kms\ $\leq$ 42.2 for \tCO, and
34.7 $\leq$ \Vlsr /\kms\ $\leq$ 44.5 for \CeO.
All the other symbols are the same as those in Fig.~\ref{fig:contmaps}.
The emission seen in the top of the upper-panels is an
artifact due to cleaning problems.
              }
         \label{fig:COmaps}
   \end{figure*}

\subsection{\tCO\ (3--2), \CeO\ (3--2), \HtCOp\ (1--0) and SiO (2--1) Line Emission}
\subsubsection{Maps of \tCO\ (3--2) and \CeO\ (3--2) Line Emission}
\label{ss:r_co}
Fig.~\ref{fig:COmaps} presents total integrated intensity maps 
of the \tCO\ and \CeO\ $J=$3--2 lines made with natural and uniform 
weightings.
For the sake of comparison,
we plot the 7\gs\ contour level
of the 6\,cm continuum emission map, i.e.
the lowest level from Fig.~\ref{fig:contmaps}.
The natural weighted maps show that 
the \tCO\ and \CeO\ (3--2) emitting regions are compact and
centered on the HMC, while these lines are not detected towards SMA3.
Since single-dish observations of similar objects indicate that
these CO isotopomers trace extended regions, 
it is reasonable to argue that in \gn\ the extended emission is
filtered out by the interferometer. We also note
that the \tCO\ and \CeO\ (3--2) lines are not seen towards SMA1 in the
uniform weighted maps.\par

\subsubsection{Spectra at the Peak Position of the HMC}
\label{sss:hmcsp}

\begin{figure}[t!]
   \centering
   \includegraphics[width=8.4cm,angle=0]{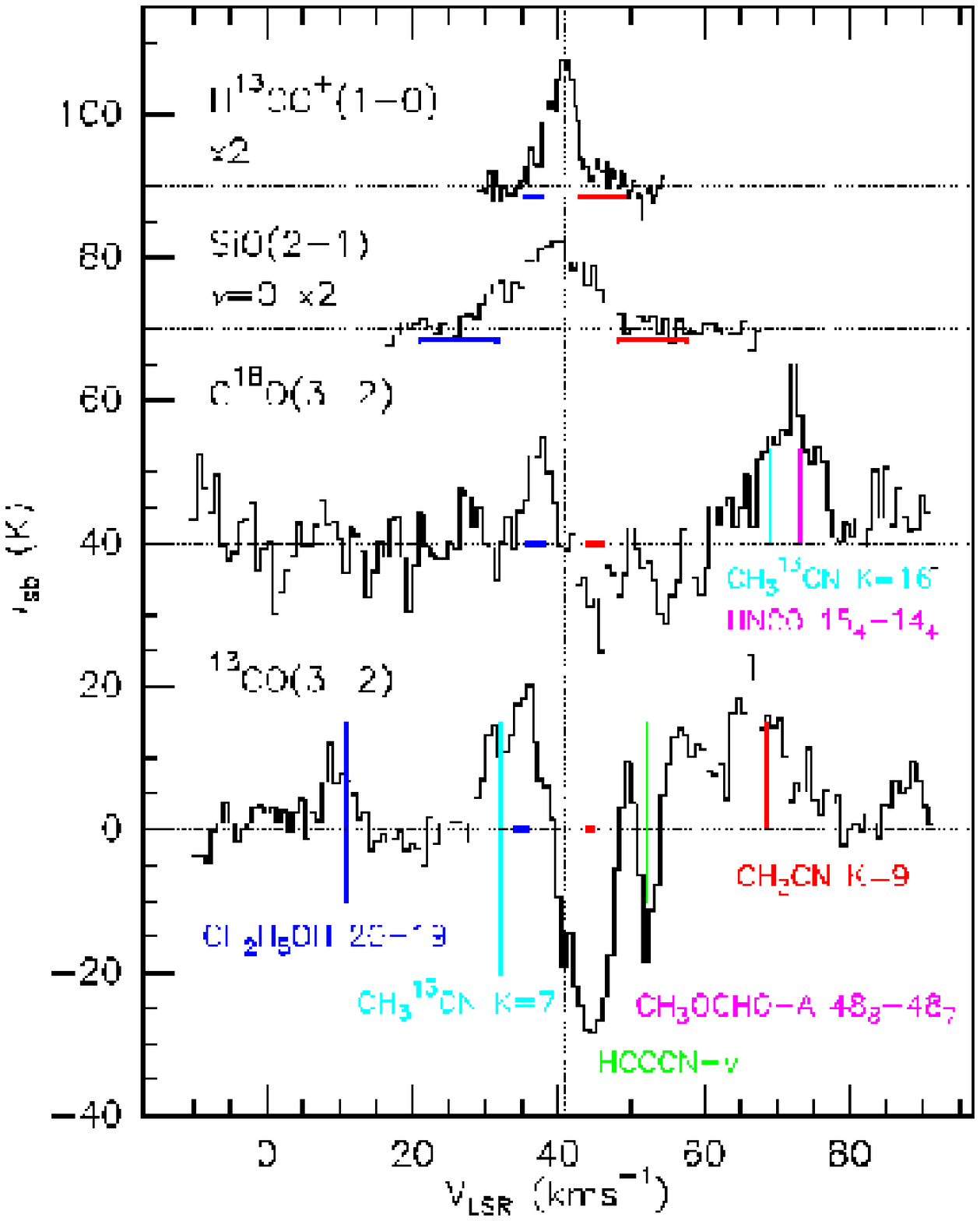} 
      \caption{
Spectra of four molecular lines in \Tsb\ scale
towards the peak of the 890 \um\ continuum emission, i.e., SMA1.
The SiO and \HtCOp\ intensities are scaled by a factor 2.0 in the \Tsb\ scale.
The vertical dashed line indicates the systemic velocity (\Vsys ) of the
cloud, \Vlsr $=$\,41.6\,\kms.
In the CO isotopomers spectra, we mark the positions of all
possible (molecular) lines, calculated assuming that they are emitted
with velocity equal to \Vsys\ (see Sect.~\ref{sss:hmcsp}).
The thick horizontal blue and red bars denote the velocity intervals over
which the line emission has been integrated to produce the maps in
Figs.~\ref{fig:HtCOp}, ~\ref{fig:HVoverlaymaps}, and ~\ref{fig:COabsmaps}.
The emission seen around \Vlsr\ $\simeq$ 70 \kms\ of the \CeO\ spectrum
is likely due only to the HNCO line.
              }
         \label{fig:SPhv}
   \end{figure}

Fig.~\ref{fig:SPhv} shows the molecular line spectra obtained from the
interferometric observations towards the peak position of SMA1.
The \HtCOp\ line peaks at \Vsys, but its line profile is not a single Gaussian.
The SiO line shows prominent high velocity wing emission.
Noticeably, the \tCO\ (3--2) and \CeO\ (3--2) lines do not show 
prominent wing emission, but present deep absorption features.
In both CO isotopomers redshifted absorption (i.e. an inverse P-Cygni
profile) is seen, albeit fainter in the \CeO\ line.
The straightforward interpretation of these finding is that the HMC is
undergoing infall.
The LSR-velocity of the deepest absorption channel is 
\Vlsr = 44.5 \kms\ for \tCO, 
and 46.0 \kms\ for \CeO,
the latter being
redshifted by 4.4 \kms\ with respect to \Vsys.
The deepest absorption channels of the \tCO\ and \CeO\ lines have 
$-29.3$\,K and $-15.0$\,K in \Tsb, respectively.
Note that the brightness temperature of the \tCO\ line (and even more so for
the \CeO\ line) is, in absolute value, less than that of the continuum peak
(see Sect.~\ref{sss:smacont}); this indicates that the absorption line is
real and not an artifact due to resolving out the extended emission.\par

Finally, we note that we cannot exclude that part of the blueshifted wing
emission of the \tCO\ (3--2) line might be due to the \mcniso\ $K=7$ line,
as we will argue in Sect.~\ref{ss:mcn}.
After considering all the other molecular lines that could overlap with the
two CO isotopomer lines, we conclude that line-contamination of the
\tCO\ line over the velocity range between \Vlsr\ $\sim$ 34 \kms\ and 50
\kms, and of the \CeO\ line over the range from $\sim$ 0 \kms\ to 65 \kms, is
unlikely.\par

%

\subsubsection{Comparisons of the Absorption Features Seen in the \tCO\ (3--2) 
and \hcop\ (1--0) Lines}
\label{sss:SPabsexam}

\begin{figure}
   \centering
   \includegraphics[width=7.2cm,angle=-90]{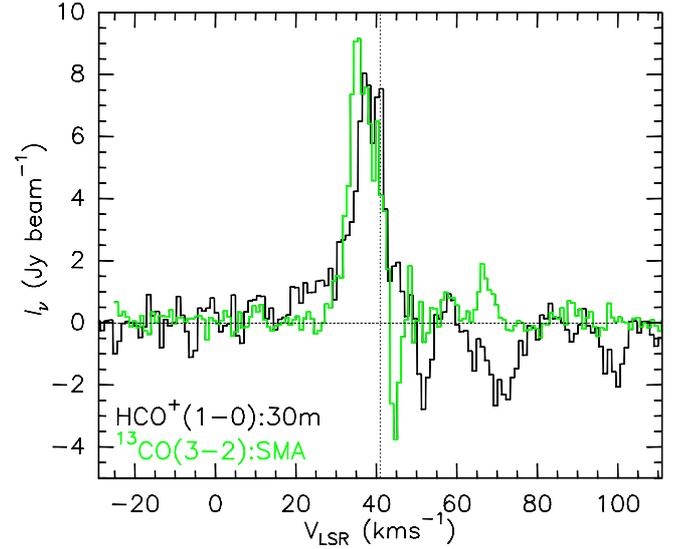} 
      \caption{Comparison of the absorption features seen in the \tCO\
    (3--2) and \HCO\ (1--0) lines toward the peak of the 890 \um\
continuum emission.
For the sake of comparison, the interferometric \tCO\ (3--2) spectrum has been
    obtained after reconstructing the data with a synthesized beam equal to the
    single-dish beam of the \HCO\ line observations with IRAM 30-m telescope
(see Sect.~\ref{sss:SPabsexam}).
              }
         \label{fig:sp_abs}
   \end{figure}

To assess the origin of the redshifted absorption seen in the
\tCO\ and \CeO\ (3--2) spectra, 
we made Fig.~\ref{fig:sp_abs} where we compare
the \tCO (3--2) spectrum with an \hcop\ (1--0) spectrum taken with
the IRAM 30\,m telescope (R. Cesaroni, unpublished data).
The \hcop\ spectrum shows a number of prominent absorption features.
To make the comparison as consistent as possible,
we reconstructed the \tCO\ image from the visibility data
adopting the same beam as the \hcop\ line
(\thetahpbw = 29\arcsec). We then took the \tCO\ spectrum towards the same
position observed in the \hcop\ line.
Notice that all the absorption features seen in the two lines are redshifted
with respect to \Vsys.
However, absorption is detected at different velocities in the two tracers:
the absorption dips are seen at
$V_{\rm LSR} - V_{\rm sys} = +3.5$ \kms\ for \tCO, 
and at $V_{\rm LSR} - V_{\rm sys}\simeq +10$, +29,
and +57~\kms\ for \hcop.
As for the two features seen at the highest velocities in \hcop,
we believe that these are due to clouds along the 
line-of-sight, because their velocities
correspond to those of the 21\,cm HI lines observed by
Kolpak et al. (2003) -- see their Fig.~4. This suggests that all the \hcop\
absorption likely occurs against the UC \hii\ regions in \gn.
In contrast, no \hcop\ absorption is detected in the velocity range
where \tCO\ and \CeO\ absorption is seen.
These facts suggest that 
the latter is due to the HMC, i.e. has
a local origin, and is thus indicating that the core is undergoing infall.
The nature of the absorption will be further discussed in Sect.~\ref{sss:abs}, 
where we study the gas infall in the core.

%

%
   \begin{figure*}
   \centering
   \includegraphics[width=14cm,angle=0]{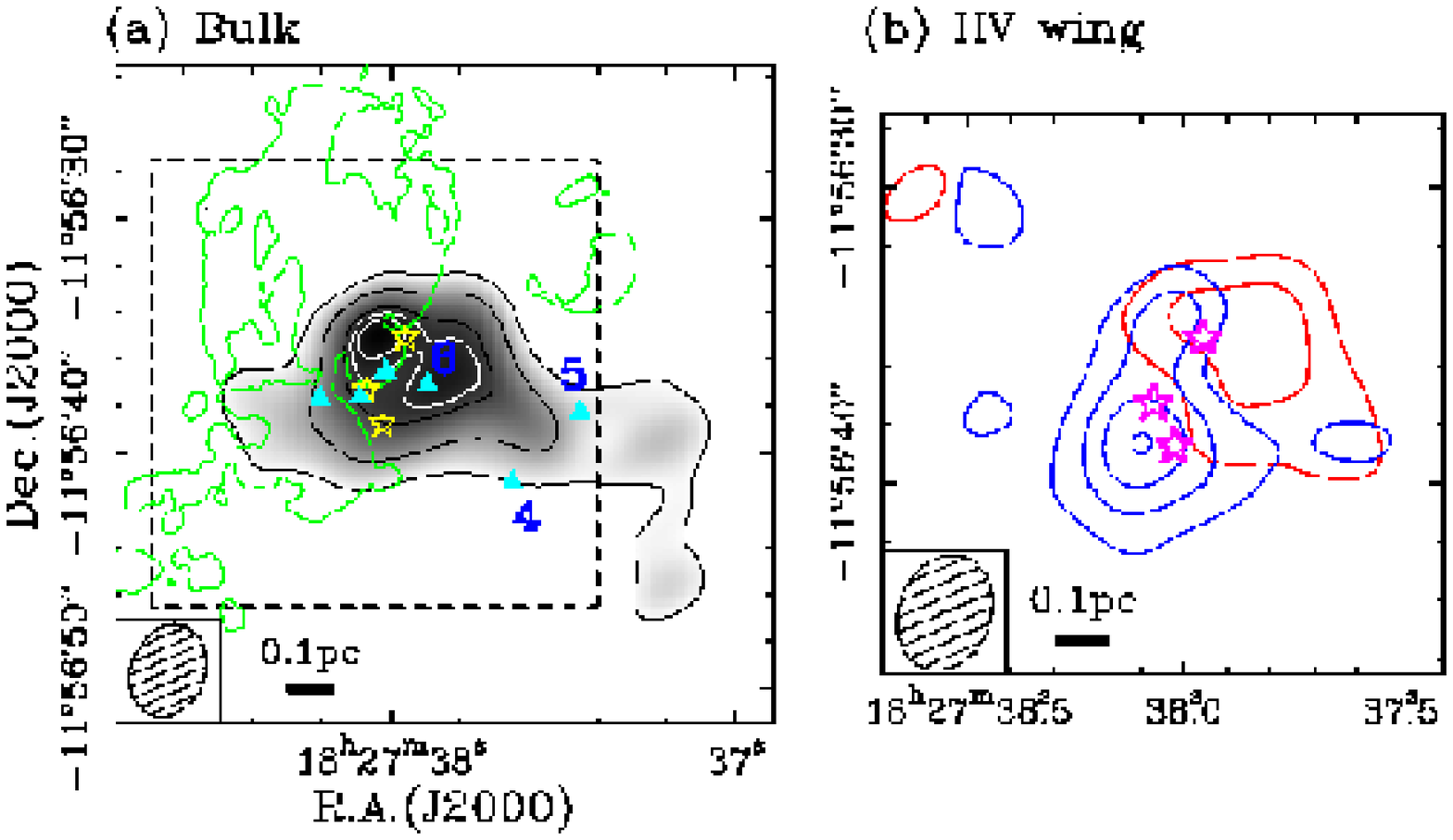} 
      \caption{(a) Integrated intensity map of the \HtCOp\ (1--0) bulk emission.
The velocity range used for the integration is 35.5 $\leq$ \Vlsr /\kms\ $\leq$ 48.9.
The interval between the thin-black contours is 2\sgm\ with the lowest contour
corresponding to the 3\sgm\ level, where the 1\sgm\ RMS noise is 0.53 Jy \pbeam\ \kms.
The thin-white contours are the 95\% and 90\% levels of
the corresponding peak intensities, where the 90\% levels correspond to the 7.5\sgm\ level.
All the other symbols are the same as those in Fig.~\ref{fig:COmaps}, and
the numbers associated with the three maser spots to the west are the same
used by Hofner \& Churchwell (1996).
The dashed box indicates the area shown in the right-hand panel. 
(b) Overlay of the blue- and redshifted wing emission maps of the \HtCOp\ (1--0) emission.
The contours are 2\sgm\ steps starting from the 3\sgm\ level.
The blue- and redshifted wing emission maps were obtained by averaging
the wing emission over the intervals
35.3 $<$ \Vlsr/\kms $<$ 37.9 and 43.1 $<$ \Vlsr/\kms $<$ 49.2, and
their RMS noise levels are 43.8 and 30.2 mJy \pbeam, respectively.
              }
         \label{fig:HtCOp}
   \end{figure*}

\subsubsection{\HtCOp\ (1--0) Emission}
\label{sss:HtCOp}

Here, we present the results of the \HtCOp\ (1--0) line
observations made with the OVRO array (Table \ref{tbl:lineobs}).
Fig.~\ref{fig:HtCOp} presents maps of both the bulk and line wing emission.
One can see that, while the former outlines a structure elongated approximately
NE--SW, the latter can be interpreted as a bipolar outflow oriented SE--NW.

Comparison with the positions of the three submm continuum sources shows
that SMA1 is the closest to the geometrical center of the outflow and is
hence the most likely candidate for powering it. 
Here the geometrical center is defined as the middle point on the line
connecting the peaks of the blue- and redshifted lobes.
We obtain (projected) angular distances to the center of 3\farcs2, for SMA1,
and 3\farcs8, for SMA2.
That SMA1 is at the origin of the flow is also supported
by comparison with the H$_2$O maser jet associated with SMA1, 
discussed in Sect.~\ref{ss:maser}.
This jet is also oriented parallel to the
\HtCOp\ outflow and has the same red--blue symmetry, which strongly suggests
a common origin for the two.\par

Using the bulk emission map, we have estimated the \HtCOp\ abundance relative to
H$_2$ ($X_{\rm H^{13}CO^+}$)
from the ratio
between the mean \HtCOp\ column density over the 3$\sigma$ contour level of
the 890~$\mu$m continuum emission (see Fig.~\ref{fig:HtCOp}a) and the
corresponding H$_2$ column density obtained from the submm continuum emission.
Such an estimate is very sensitive to the temperature of the gas and dust. 
In our calculation, we have assumed that the gas and dust are 
well-coupled, thus having identical excitation and
dust temperatures. 
For a fiducial value of 80~K (Sect.~\ref{sss:smacont}), we obtain an abundance of
$\sim10^{-10}$, but one should keep in mind that $X_{\rm H^{13}CO^+}$
spans a range from $\sim3~10^{-11}$ to $\sim6~10^{-10}$
for $T$ varying by a factor 2 with respect to the fiducial value.\par

Assuming $X_{\rm H^{13}CO^+}=10^{-10}$ and $T=80$~K (Sect.~\ref{sss:smacont}), 
we have computed the
outflow parameters by integrating the line emission 
for the blue and red wings (see caption of Fig.~\ref{fig:HtCOp} for their velocity ranges).
With these assumptions, we obtain a total outflow mass (\Mlobe) of
3700~\Msun\ and a momentum of 14000~\Msun\,\kms. Note that, even ignoring the
error on $X_{\rm H^{13}CO^+}$, a factor 2 uncertainty on the gas temperature
affects by an additional factor $\sim$2 the outflow parameters.
The dynamical timescale (\td) is estimated from the ratio between the diameter of
the lobes and the difference (in absolute value) between the systemic
velocity and the terminal wing velocity.
We obtain $\sim 8\times10^4$~yr, without correcting for the (unknown) outflow
inclination.  The mass loss rate ($\dot{M}_{\rm flow}$) and 
a momentum rate  ($F_{\rm flow}$) are thus
0.05~\Msun\,yr$^{-1}$ and 0.17~\Msun\,\kms\,yr$^{-1}$, respectively. Values
that large are very likely overestimated due to the various assumptions
made.  Nevertheless, they indicate that the powering source should be as
luminous as $\sim10^5$~\Lsun, according to the empirical relationship derived
by Beuther et al.~(2002). Here, we have implicitly assumed that one is
dealing with a single star. If this is correct, then such a star must be in a
pre-\hii\ region phase, as no free-free continuum emission has been detected
towards the HMC. 
We will further discuss this in Sect.~\ref{ss:evolution} and Sect.~\ref{ss:SMA1nature}.

   \begin{figure*}
   \centering
   \includegraphics[angle=0,width=14cm]{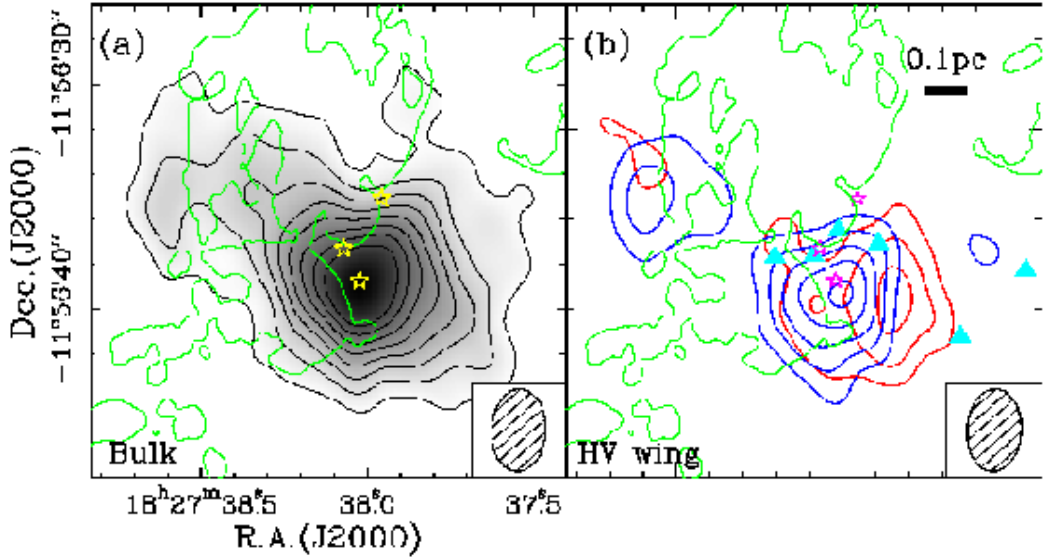} 
   \caption{(a) 
Integrated intensity map of the bulk emission of the SiO (2--1) $v=0$ line
(greyscale plus thin contours) observed with the OVRO and NMA interferometers.
The green contours have the same meaning as in Fig.~\ref{fig:COmaps}.
The bulk emission has been integrated over the LSR-velocity range of 
31.9 $\leq$ \Vlsr /\kms\ $\leq$ 48.4, 
and the RMS noise level is 0.28 Jy \pbeam\ \kms.
The yellow stars
represent the positions of SMA1, SMA2, and SMA3 (see Table \ref{tbl:smacont}).
(b) Overlay of the blue- and redshifted wing emission maps of the
SiO (2--1) $v=0$ emission.
The contours are spaced by 2\sgm\ and start at the 3\sgm\ level.
The blue- and redshifted wing emission maps have been obtained by averaging
the wing emission over the intervals
21.1 $<$ \Vlsr/\kms $<$ 31.9 and 48.4 $<$ \Vlsr/\kms $<$ 58.0, and
their RMS noise levels are  11.6 and 13.9 mJy \pbeam, respectively.
The hatched ellipses in the bottom right indicates
the FWHM of the synthesized beam (Table \ref{tbl:lineobs}).
}
   \label{fig:HVoverlaymaps}
   \end{figure*}

\subsubsection{SiO (2--1) $v=0$ Emission}
\label{ss:sio}

\paragraph{Maps of the SiO Emission --- }

From Fig.~\ref{fig:SPhv} one sees that
the \Vt\ of the SiO emission is blueshifted by 
$\sim 11$ \kms\ and redshifted by $\sim 15$ \kms\ with respect to \Vsys.
The presence of HV wing emission strongly suggests the existence of
a molecular outflow driven by a YSO in one of the cores.
In order to analyze the structure of the SiO emitting gas, 
we produced maps of the bulk emission and HV wing emission 
in Fig.~\ref{fig:HVoverlaymaps}.
In panel (b), one sees that the HV wing emission is
elongated along the east-west direction, with the blue lobe lying to east and
the red lobe to the west.  Such a bipolarity indicates that the SiO HV gas
very likely traces a bipolar outflow, although the lobes do not appear very
collimated.  The fact that the two lobes are largely overlapping each other
suggests that the outflow is seen close to pole-on. Interestingly, the velocity
structure of the SiO outflow is very different from that of the larger scale
($\sim$20\arcsec) \tCO\ (2--1) outflow mapped by Lo\'pez-Sepulcre et al.
(2009) with the IRAM 30-m telescope. The latter has the redshifted lobe to
the NE and the blueshifted to the SW. We stress that this is not due to the
different velocity intervals used to produce the outflow maps, because the
orientation of the SiO outflow does not change using the same velocity ranges
adopted by Lo\'pez-Sepulcre et al. (narrower than those used by us).
We conclude that in all likelihood the larger scale flow is originating from
another YSO in the region.\par

What is the source powering the SiO outflow? We believe that it is SMA2,
despite the small offset between this and the geometrical center of the
outflow (see Fig.~\ref{fig:HVoverlaymaps}b), because
Fig.~\ref{fig:HVoverlaymaps}a shows that the peak of the SiO bulk
emission is clearly coincident with SMA2 
(see Fig.~\ref{fig:HVoverlaymaps}a).\par

Note that this implies that (at least) three bipolar outflows are present in
the region. Beside the large-scale flow mapped by Lo\'pez-Sepulcre et al.,
we have detected two compact outflows: one traced by \HtCOp\
(see Sect.~\ref{sss:HtCOp}) and powered by a YSO in SMA1,
and another traced by SiO and powered by a YSO embedded in SMA2.
Is the latter as massive as the former? 
Next,  we attempt to estimate luminosity of the YSO from the SiO outflow parameters.

\paragraph{Properties of the SiO Outflow driven by SMA2 --- }
To derive the mass and kinematical parameters of the outflow, we assume that
the SiO line is optically thin and the SiO molecule is in LTE.  We adopt an
excitation temperature (\Tex ) of 20\,K (Appendix \ref{as:pd}) and an
SiO/H$_2$ abundance ratio of $3\times10^{-9}$.  The latter is estimated from
the ratio between the SiO column density (obtained from the bulk emission --
Fig.~\ref{fig:HVoverlaymaps}a) and the H$_2$ column density (calculated from
the 890 \um\ continuum emission towards SMA\,2 -- Table \ref{tbl:contobs}).
We note that that the outflow masses are likely underestimated because of
missing-flux filtered out by our interferometric observations and the
uncertainty in defining a boundary velocity (\Vb ) between the outflowing gas
and the quiescent ambient gas.\par

We find an outflow mass of 90~\Msun\ and a momentum of 2100~\Msun\,\kms.  
We also obtain \td\ $\simeq$ $2 ~10^4$\,years, implying an outflow mass-loss rate
$\dot{M}_{\rm flow} \simeq 5\times10^{-3}$ \Msun\ yr$^{-1}$ and a momentum rate 
$F_{\rm flow}\simeq$\,0.06~\Msun\,\kms\,yr$^{-1}$.
Since outflows are believed to be momentum driven (\cite{cb92}), the latter
value may be taken as an indicator of the outflow strength and hence of the
mass and luminosity of the YSO powering it.
Using the relationship between $F_{\rm flow}$ and YSO luminosity
obtained by Beuther et al. (2002),
we estimate that the YSO powering the SiO outflow should be as luminous as
$\sim3\times10^4$~\Lsun. 
This indicates that the powering source is a high-mass star.\par

When deriving the outflow parameters as done here with SiO and in
Sect.~\ref{sss:HtCOp} with \HtCOp, a word of caution is in order.
The difference between the masses of the \HtCOp\ outflow driven by SMA1
and the SiO outflow from SMA2 amounts to a factor $\sim$40. Such a large
number may be due to multiple outflows unresolved in our observations
as well as to the uncertainties on the fractional abundances of the
two molecules, which are difficult to predict. These caveats cast some doubt on the
values derived in our calculations. Therefore, although the basic conclusion
that the two outflows are associated with high-mass YSOs is likely to be
correct, the outflow parameters reported here must be considered
with caution.\par

%
   \begin{figure*}[t]
   \centering
   \includegraphics[width=11cm,angle=0]{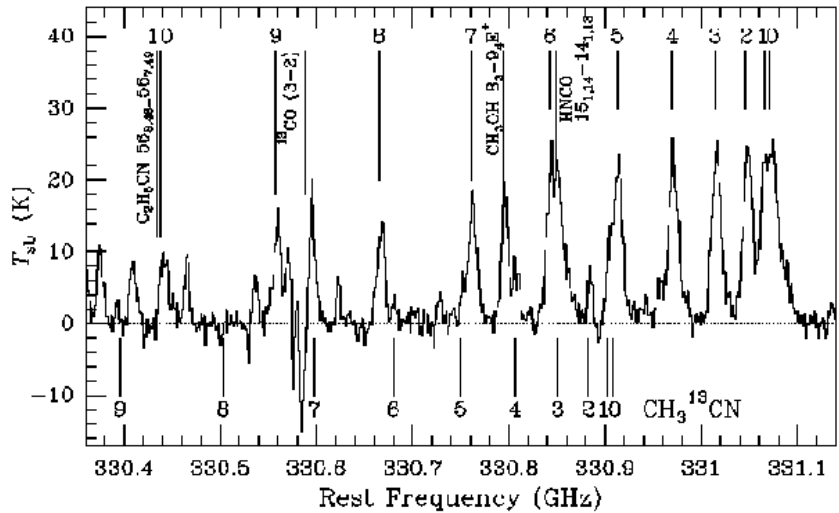}
      \caption{Interferometric spectrum of the \mcn\ and \MCNII\ (18--17) lines towards
the peak position of the 890 \um\ continuum source, SMA1.
The vertical bars above and below the spectrum indicate the rest-frequencies
of the $K$ components
of the \mcn\ and \MCNII\ $J=18-17$ transitions.
We also indicate other lines that may be blended with the
methyl cyanide lines.
              }
         \label{fig:SPmcn}
   \end{figure*}
%
  \begin{figure*}
   \centering
   \includegraphics[width=12cm,angle=0]{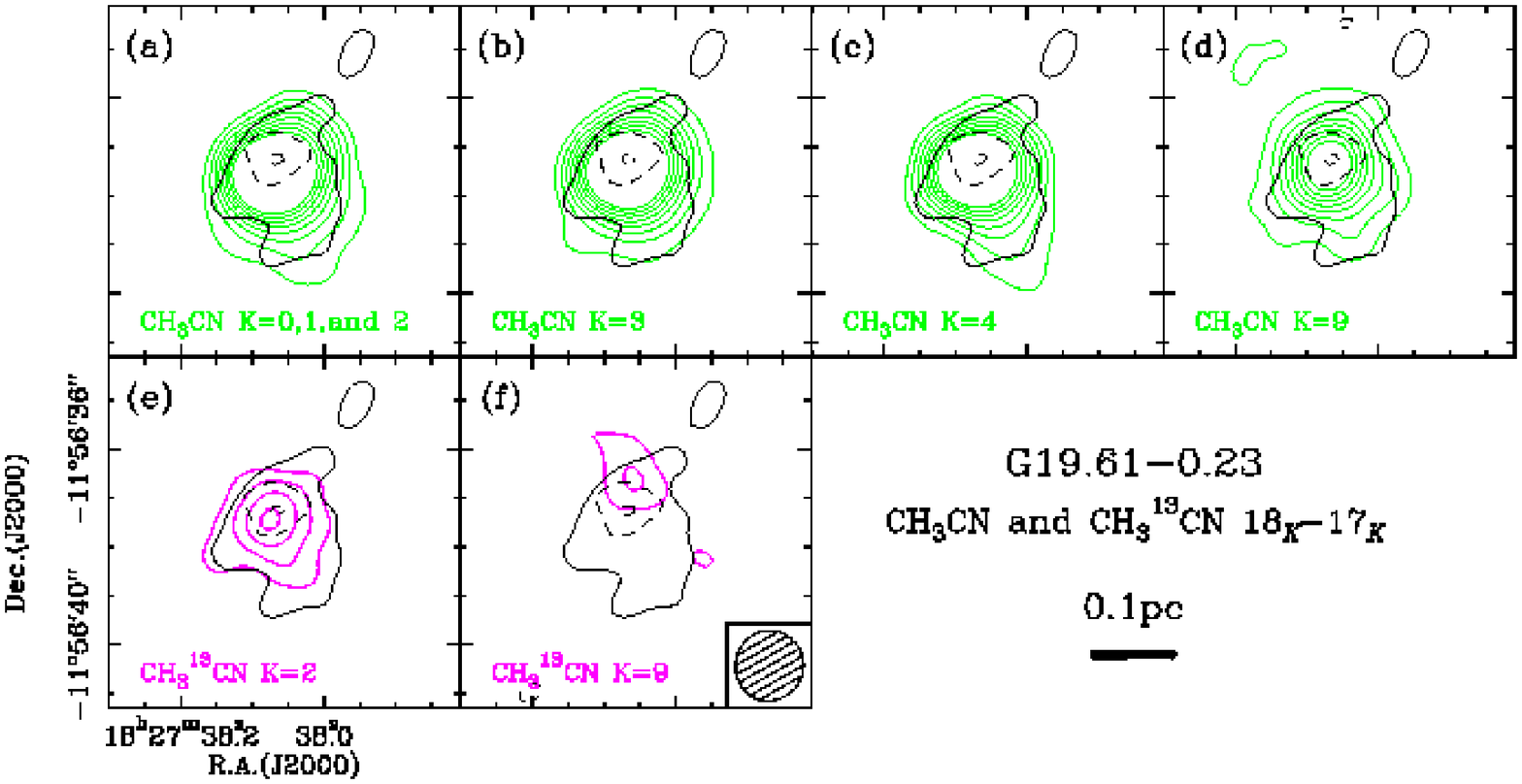} 
      \caption{Integrated intensity maps of the $J_K=18_K-17_K$ 
emission of the \mcn\ (green contour) and \MCNII\ (magenta contour) lines.
The black dashed contour corresponds to the absorbtions region seen in the \tCO\ (3--2) line.
The thin black contour represents the 7\sgm\ level of the uniform weighting
890 \um\ continuum emission map (Fig.~\ref{fig:smacont}b).
All the transitions, except the \mcn\
    $K=$\,0,1, and 2 components, are not blended with other lines
(see Fig.~\ref{fig:SPmcn} for the spectrum). 
The contour intervals are spaced by 2\sgm\ and start from the
3\sgm\ level.
In the \mcn\ maps, we do not plot contours above the 17\sgm\ level
to prevent saturation of the maps.
The 1\sgm\ RMS noise levels are 6.3, 1.4, 1.3, 0.33, and 0.17 mJy \pbeam\ \kms\
for the \mcn\ (18--17) $K=0$ to 2 emission, 
$K=3$, $K=4$, and \MCNII\ (18--17) $K=2$, and $K=6$, respectively.
              }
         \label{fig:MCNmaps}
   \end{figure*}

\subsection{\mcn\ $18_K-17_K$ Emission}
\label{ss:mcn}
\subsubsection{Spectrum}
\label{sss:mcnsp}

Fig.~\ref{fig:SPmcn} shows the spectrum of the \mcn\ and \mcniso\ $J_K =
18_K-17_K$ lines observed with the SMA towards the 890 \um\ continuum peak,
SMA1 (Sect.~\ref{sss:smacont}).  High-excitation $K$ components can be seen in
this spectrum. In particular, one can recognize all lines up to $K=$\,10
for \mcn, and up to $K=$\,6 for \mcniso,
%
%
although overlap with transitions from other species (some
of which are marked in the figure) cannot be excluded.
Only the \mcn\ (18--17) $K=3$ and $K=4$
and  the \mcniso\ (18--17) $K=2$ and 6 components do not appear to be
affected by blending.

The peak intensities of the $K=0$ to 4 lines of the \mcn\ are comparable
each other,
suggesting that these \mcn\ lines are optically thick.  One can calculate the
optical depth of the corresponding \mcniso\ lines from the ratio between the
two isotopomers. The values of $\tau$ for the $^{13}$C substituted species
range from 0.1 to 0.5 for $K\le5$. These imply optical depths of 5 to 29 for
\mcn, assuming a $^{12}$C/$^{13}$C abundance ratio of 55 at a
galactocentric distance of 5.3~kpc (\cite{wiro}). As we will discuss in
Sect.\ref{sss:mcn_excondition}, opacities that large hinder the usage of
rotation diagrams to estimate the gas temperature and column density.

%
   \begin{figure}[t!]
   \centering
   \includegraphics[width=5.8cm,angle=-90]{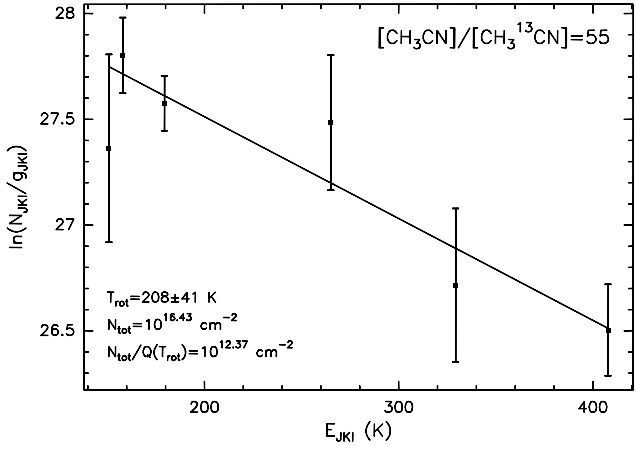} 
      \caption{Rotation diagram obtained from the $K=$\,0, 1, 2, 4, 5,  {\rm and} 6
      components of the \mcniso\ (18--17) lines shown in Fig.~\ref{fig:SPmcn}.
              }
         \label{fmcn}
   \end{figure}

\subsubsection{Maps: Comparison with the Other Lines}
\label{sss:MCNmaps}

Fig.~\ref{fig:MCNmaps} shows maps of some \mcn\ and \MCNII\ (18--17) lines
that are not blended with transitions of other molecules.
For the sake of comparison, in the same figure we also outline
the uniform weighted
890\,\um\ continuum map (Fig.~\ref{fig:smacont}b) 
and map of the redshifted absorption seen in the \tCO\ (3--2) emission,
described in Sect.~\ref{sss:hmcsp}.
The \tCO\ (3--2) map corresponds to
the velocity channel where the maximum absorption is attained,
i.e., at \Vlsr\ $=$ 44.0 \kms.
We remind that no emission is seen towards SMA1 in the integrated intensity
maps of the \tCO\ and \CeO\ (3--2) lines  (Fig.~\ref{fig:COmaps}).
From all this, one obtains the following results:
(i) \mcn\ and \mcniso\ line emission is
detected towards SMA1 and SMA2, but not towards SMA3;
(ii) the peak of the methyl cyanide emission
coincides with the peak of SMA1;
(iii) the redshifted \tCO\ absorption is seen towards the center
of the HMC (i.e., SMA1), but is not detected
towards SMA2 and SMA3;
(iv) the emitting region of the low $K$-transition, e.g., $K=0$, 1, and 2,
has a size similar to that of the the higher $K$-lines, while in
similar objects (e.g., Beltr\'an et al. 2004) the emitting region is
smaller for higher excitation transitions.

\subsubsection{Excitation Conditions of \mcn}
\label{sss:mcn_excondition}

Methyl cyanide is known to be an excellent temperature tracer and rotation
diagrams obtained from the different $K$ components are commonly used to
estimate the gas temperature and column density. However, as argued in
Sect.\ref{sss:mcnsp}, most \mcn(18--17) transitions are optically thick thus
making the column density in the corresponding level a lower limit. To
circumvent this problem, we have used the optically thin ($\tau<0.5$ -- see
Sect.\ref{sss:mcnsp}) lines of \mcniso. Despite heavy blending with other
transitions, we managed to obtain an estimate of the line parameters by
fitting groups of $K$ components simultaneously, by fixing their separation in
frequency to the laboratory values and forcing the line widths to be the
same. In this way we could successfully fit the $K=0, 1, 2, 4, 5, {\rm and}~ 6$
components and produce the rotation diagram shown in Fig.~\ref{fmcn}. In our
calculation, we have assumed a source angular diameter of 
2\farcs4 from Table~ \ref{tbl:smacont}
and a $^{12}$C/$^{13}$C abundance ratio of 55. 
The best fit gives a temperature of 208$\pm$ 41~K 
and a source averaged \mcn\ column density of
$2.7\times10^{16}$~cm$^{-2}$. While the latter is consistent with 
the values quoted by Wu et al. (2009) and Qin et al. (2010), 
the former is significantly less than their estimates of 
552$\pm$29~K 
(\cite{wu09}) and 609$\pm$77~K (\cite{qin10}).
Such a discrepancy is due to these authors using the optically thick
\mcn\ lines in their calculations, which
leads to a severe overestimate of the true
rotation temperature.\par

%
   \begin{figure}[t]
   \centering
    \includegraphics[width=6.5cm,angle=0]{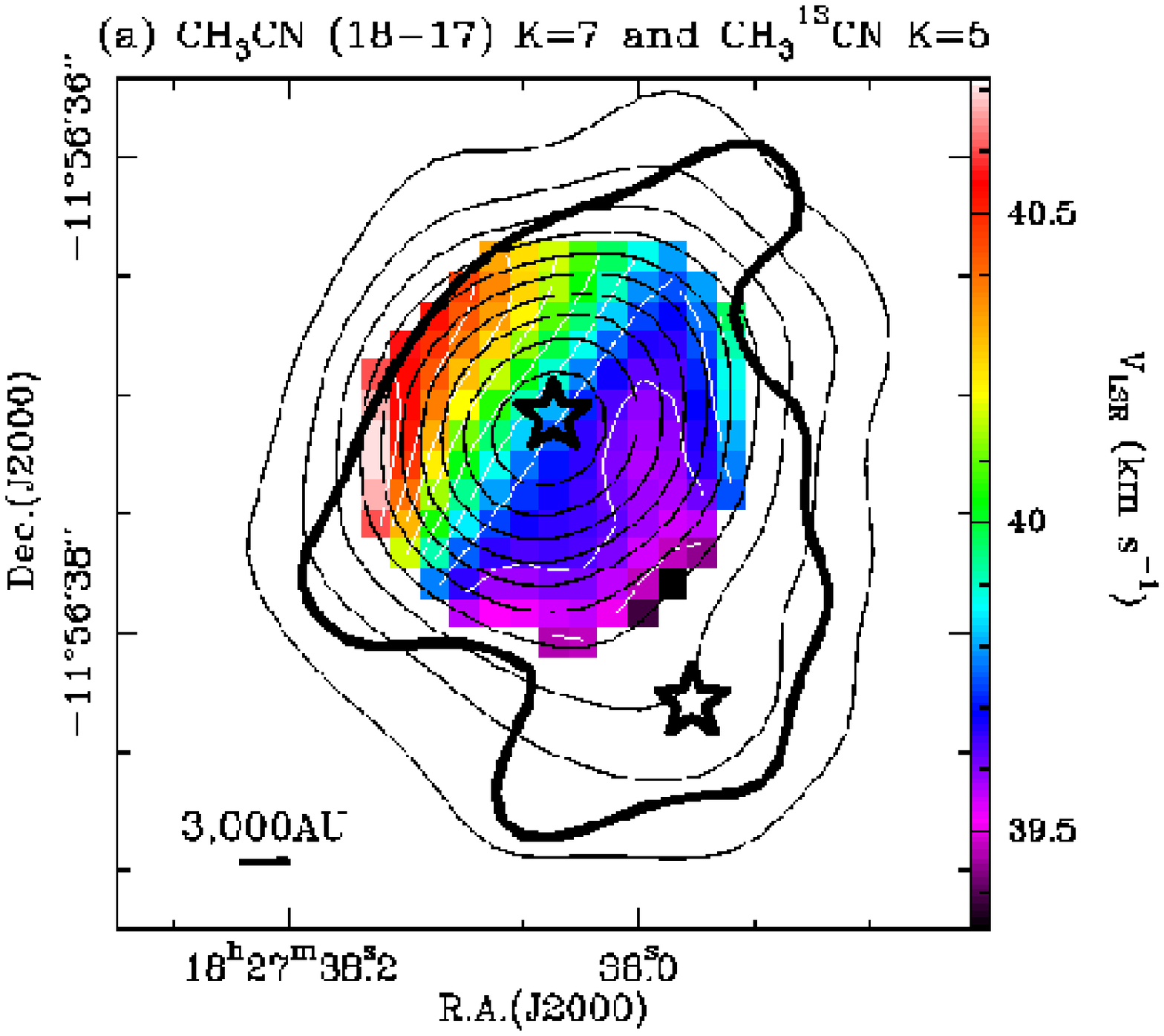} \\ 
    \includegraphics[width=6.5cm,angle=0]{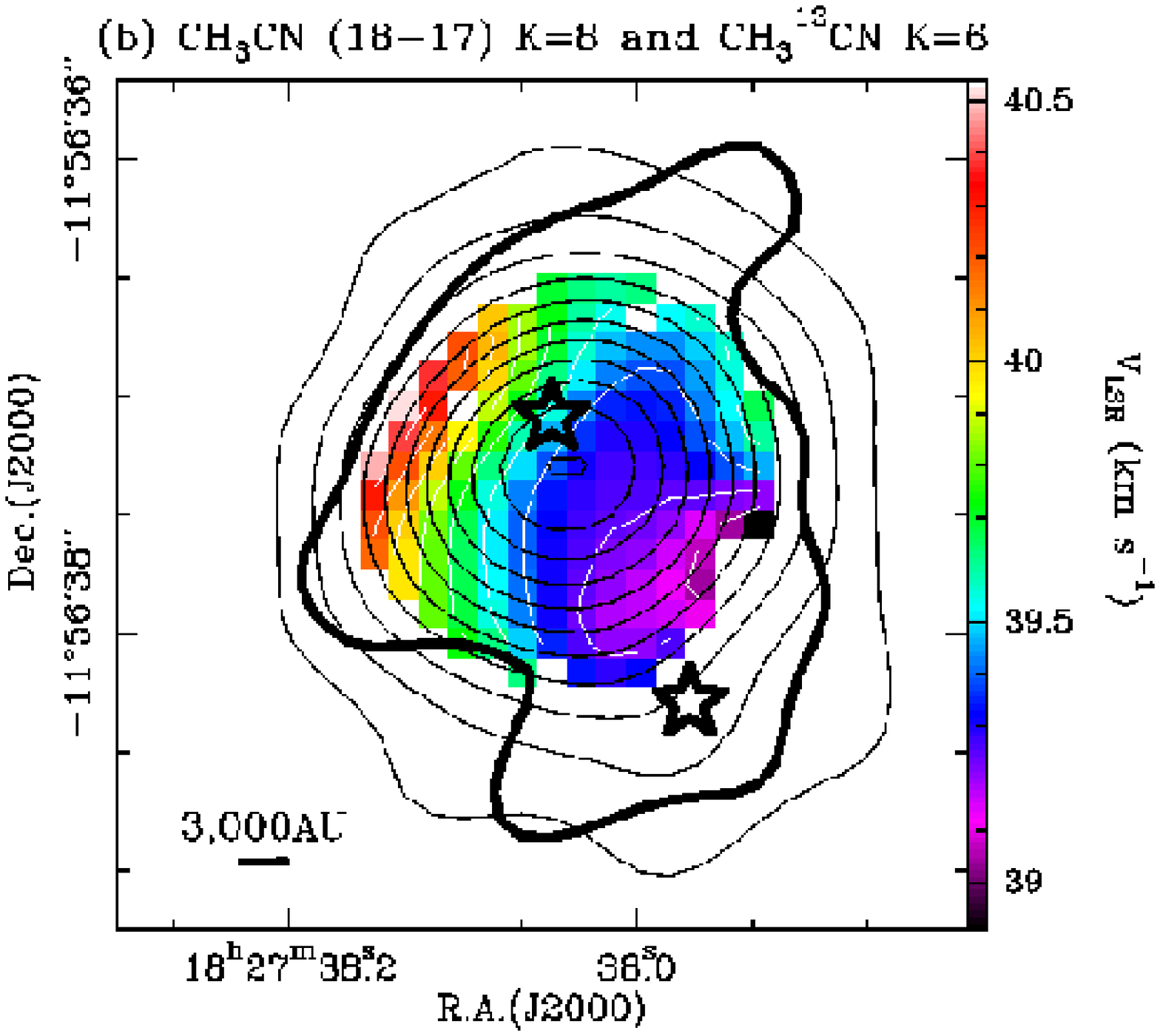}
      \caption{Maps of the line peak velocity (color) 
obtained from simultaneous Gaussian fits to the
\mcn\ and/or \mcniso\ $J_K=18_K-17_K$ lines (see Sect.~\ref{sss:Vmap} for details).
The names of the lines above each panel show the transitions that 
have been analyzed.
The thin contours are a map of the integrated emission of the corresponding lines.
The contours are in 2\sgm\ steps, starting from the 3\sgm\ level.
the thick contours corresponds to the 5\sgm\ level of the 890 \um\ continuum
emission shown in Fig.~\ref{fig:smacont}b).
The two stars mark the peak positions of SMA1 and SMA2 (Sect.~\ref{sss:smacont}).
         \label{fig:velmaps}}
   \end{figure}

Since the source angular diameter is comparable to the SMA beam and the \mcn\
line are optically thick, the line brightness temperature should be
similar to the gas kinetic temperature of $\sim$200~K. Instead, only
$\sim$20~K are measured with the SMA (see Fig.~\ref{fig:SPmcn}), implying a
beam filling factor of 0.1.  Although the existence of sub-structures due to
clumpiness on angular scales smaller than the interferometer beam is very
likely, such a filling factor appears too small, as we do not reveal
important fragmentation of the HMC in our maps. We thus believe that beam
dilution may explain only in part the low value of \Tsb\ and
conclude that the optically thick \mcn\ lines must be tracing the outer
regions of the core, where the gas temperature is significantly less than the
temperature measured in the thinner \mcniso\ transitions, originating from
the innermost regions.

\subsubsection{Velocity Structure of the HMC Traced by
\mcn\ Emission}
\label{sss:Vmap}
To investigate the velocity field of the innermost part of 
the SMA1 core, in Fig.~\ref{fig:velmaps} ,
we plot maps of the \mcn\ line centroid velocity over the HMC.
This velocity field was obtained by fitting simultaneously
multiple $K$-components with Gaussian profiles 
with separations in frequency fixed
to the laboratory values (see Pearson \& Mueller 1996)
and line widths forced to be equal.
The method employed here is described in 
e.g., Beltr\'an et al. (2005) and Furuya et al. (2008).
Considering the line-blending and high opacity at low-$K$ lines of \mcn\ 
(Sect.~\ref{sss:mcnsp}), we decided to fit the
\mcn\ $K=7$ and \mcniso\ $K=5$ lines simultaneously, to
obtain the map in Fig.~\ref{fig:velmaps}a,
and the \mcn\ $K=8$ and \mcniso\ $K=6$ lines, for the map in
Fig.~\ref{fig:velmaps}b.
The values of the velocity are displayed only inside the
area encompassed by the 9\sgm\ contour level of the corresponding
integrated emission map.\par

Unlike other cases (Beuther et al. 2005), in our study the \mcn\ lines
appear to trace a clear velocity gradient, increasing from SW to NE,
as observed in similar sources (e.g. Beltr\'an et al. 2004).
In the \gn\ HMC, the inferred velocity field maps 
show a coherent pattern,
suggestive of the presence of systematic motions of the gas.
In addition, the two maps show a fairly nice consistency.
To further confirm these findings, we made velocity-field 
maps by fitting a single-Gaussian profile to the weak, optically thin
\mcniso\ $K=2$ component,
which is not blended with other transitions.
Also in this case, the velocity map
confirms the existence of a velocity gradient and
demonstrates that this is not due to opacity effects.

Note that the velocity gradient is almost perpendicular to
the axis of the \wat\ maser jet  in Fig.~\ref{fig:masermap}
and that of the \HtCOp\ outflow in Fig.~\ref{fig:HtCOp}b.
Therefore, our interpretation of the velocity gradient is that
the HMC is rotating about the jet/outflow axis
oriented SE--NW, alike the ``toroids'' imaged by Beltr\'an et al.~(2006).

\section{Discussion}
\label{s:discussion}
\subsection{Evolutionary Stage of the Massive YSOs in the 3 Submm Sources}
\label{ss:evolution}

Previous studies (e.g., Codella et al. 1997) have shown that \wat\ masers are
preferentially coincident with dense molecular cores that do not show
continuum emission from ionizing gas, implying that such cores may harbor
massive YSOs prior to the formation of an \hii\ region. This is also
consistent with the case of low- and intermediate mass stars, where
\wat\ masers are known to be associated with the youngest evolutionary phases
(Furuya et al. 2001; 2006).  Given the large masses of the three submm cores
in \gn\ -- on the order of $10^1$--$10^3$\,\Msun\
-- one can hypothesize that each of them could develop an \hii\ region; the
fact that, instead, no free-free emission is detected inside the cores
(Sect.~\ref{sss:vlacont}), as well as the presence of a \wat\ masers jet in SMA1
(Sect.~\ref{ss:maser}), suggest that the YSOs harbored in the cores are massive
but in an early evolutionary stage.\par

It is possible to constrain the properties of the putative \hii\ regions and
corresponding ionizing stars embedded in the cores using the upper limits
obtained from our observations of the continuum emission. For this purpose,
we use the method illustrated by CTC97 (see also Molinari et al. 2000).\par

%
\begin{figure}
   \centering
   \includegraphics[width=6.4cm,angle=-90]{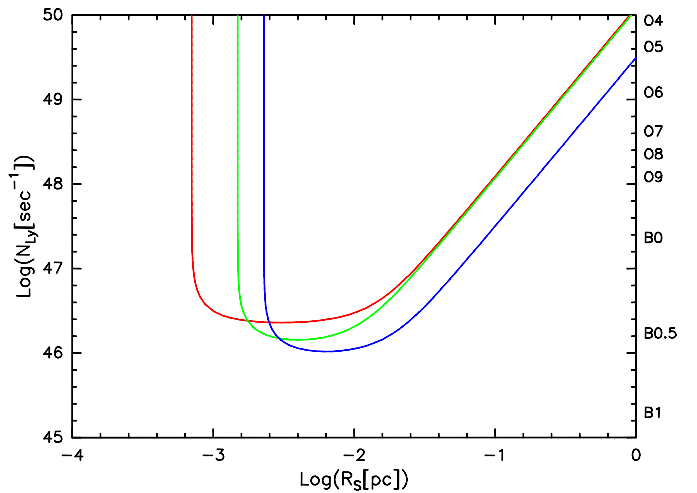}
   \caption{Plot of the brightness temperature in the synthesized beam of
a Str\"omgren \hii\ region, as a function of radius and 
Lyman continuum photon rate (see Sect.\ref{ss:evolution} for details).
The curves correspond to the 3\sgm\ upper limits on the free-free
continuum emission measured towards SMA1 at
4.86\,GHz (blue),
8.42\,GHz (green), and
22.273\,GHz (red).
Here the RMS noise levels and the geometrical mean of the beam FWHPs 
adopted to plot are
230 $\mu$Jy \pbeam\ at 5\,GHz,
180 $\mu$Jy \pbeam\ at 8\,GHz, and
360 $\mu$Jy \pbeam\ at 22\,GHz.
The labels to the right indicate the
spectral types of the ZAMS stars with the corresponding value of $N_{\rm Ly}$,
according to Panagia (1973).
}
\label{fig:hii_det}
\end{figure}

For the sake of simplicity, we arbitrarily assume that each submm source
develops a single massive YSO, instead of a cluster.  As explained in CTC97,
the peak brightness temperature, \Tsb, of a Str\"omgren \hii\ region at a
given frequency can be expressed as a function of the radius, \Rs, and 
Lyman continuum photon rate, \Nly, of the star.  
In the calculation, we also adopted
a source distance of 12.6\,kpc, and an electron temperature (\Te) of 7200\,K,
corresponding to the mean for the UC \hii\ regions in \gn\ (\cite{garay98}).
For a given \Tsb, one obtains a curve like those plotted in
Fig.~\ref{fig:hii_det}.  Note that we have not considered the 7\,mm image
(Sect.~\ref{sss:vlacont}), because this has resolution and sensitivity about 3--5
times worse than the cm images.  The three curves correspond to 3\sgm\ upper
limits obtained from the VLA continuum maps at 1.3, 3.8, and 6~cm. The
permitted values of \Tsb\ are those lying under each curve. An additional
constraint is set by the maximum radius of the \hii\ region, which cannot be
larger than the core radius, namely 0.025--0.072~pc, depending on the core
(Table~\ref{tbl:smacont}).\par

If the putative embedded stars are massive, i.e. earlier than approximately
B0.5, they must be also very young, as the corresponding \hii\ regions cannot
be larger than $6\,10^{-4}$\,pc\,=\,130~AU, which means that they are basically
quenched. On the other hand, we cannot rule out the possibility that the
stars are later than B0.5, and in this case the \hii\ regions could be larger
and optically thin. We believe, though, that the latter possibility is less
likely given the above mentioned large masses of the cores and the signposts of
high-mass star formation associated especially with SMA1 (i.e.
water masers, high temperature, and high-excitation lines),
and thus conclude that in all likelihood
the HMC hides OB type stars in a pre-UC~\hii\ region phase.

\subsection{Velocity Structure of the SMA1 Core: Infall plus Rotation}
\label{ss:SMA1VelStr}

\begin{figure}
   \centering
   \includegraphics[width=6.5cm,angle=0]{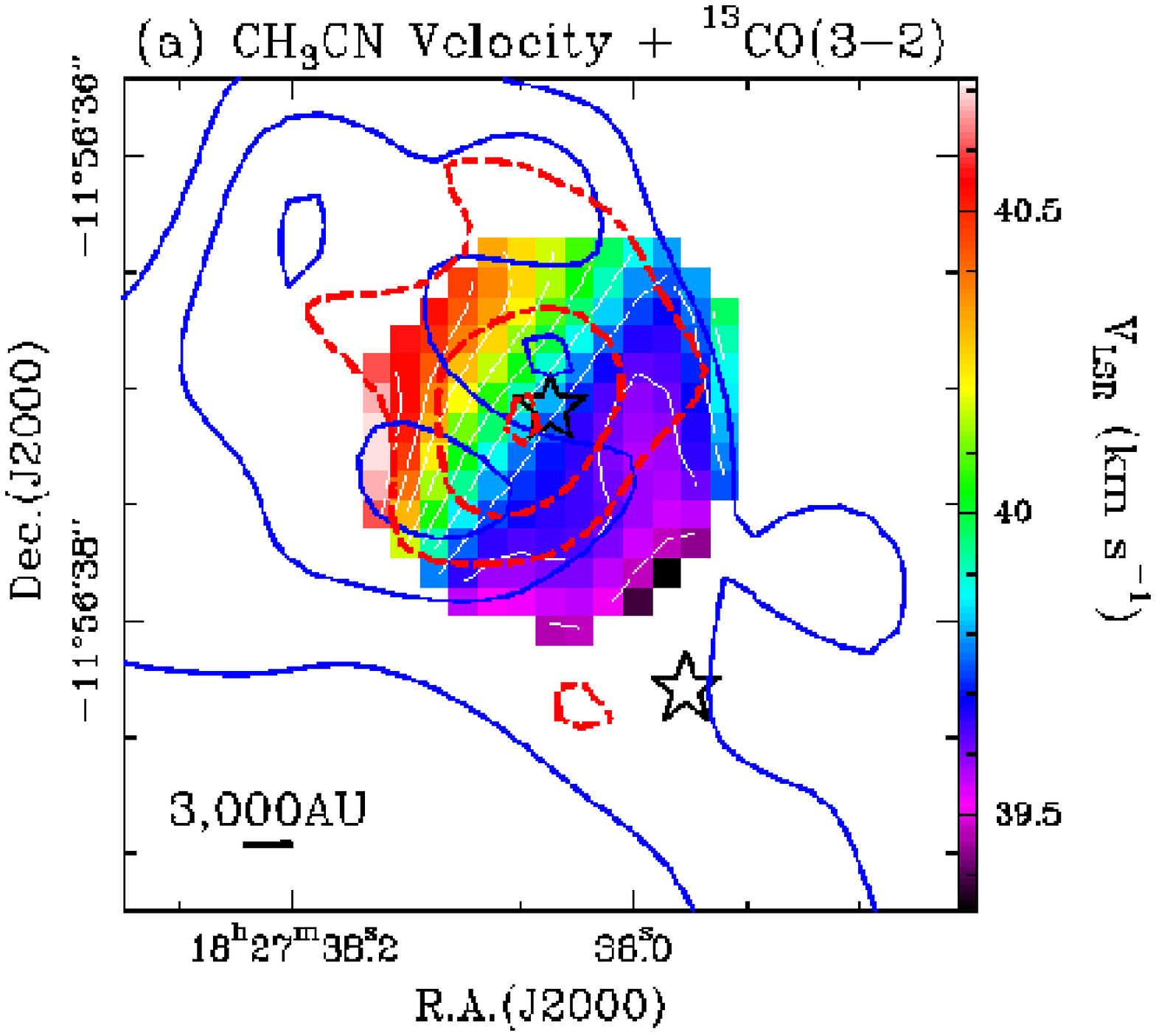} \\ 
   \includegraphics[width=6.5cm,angle=0]{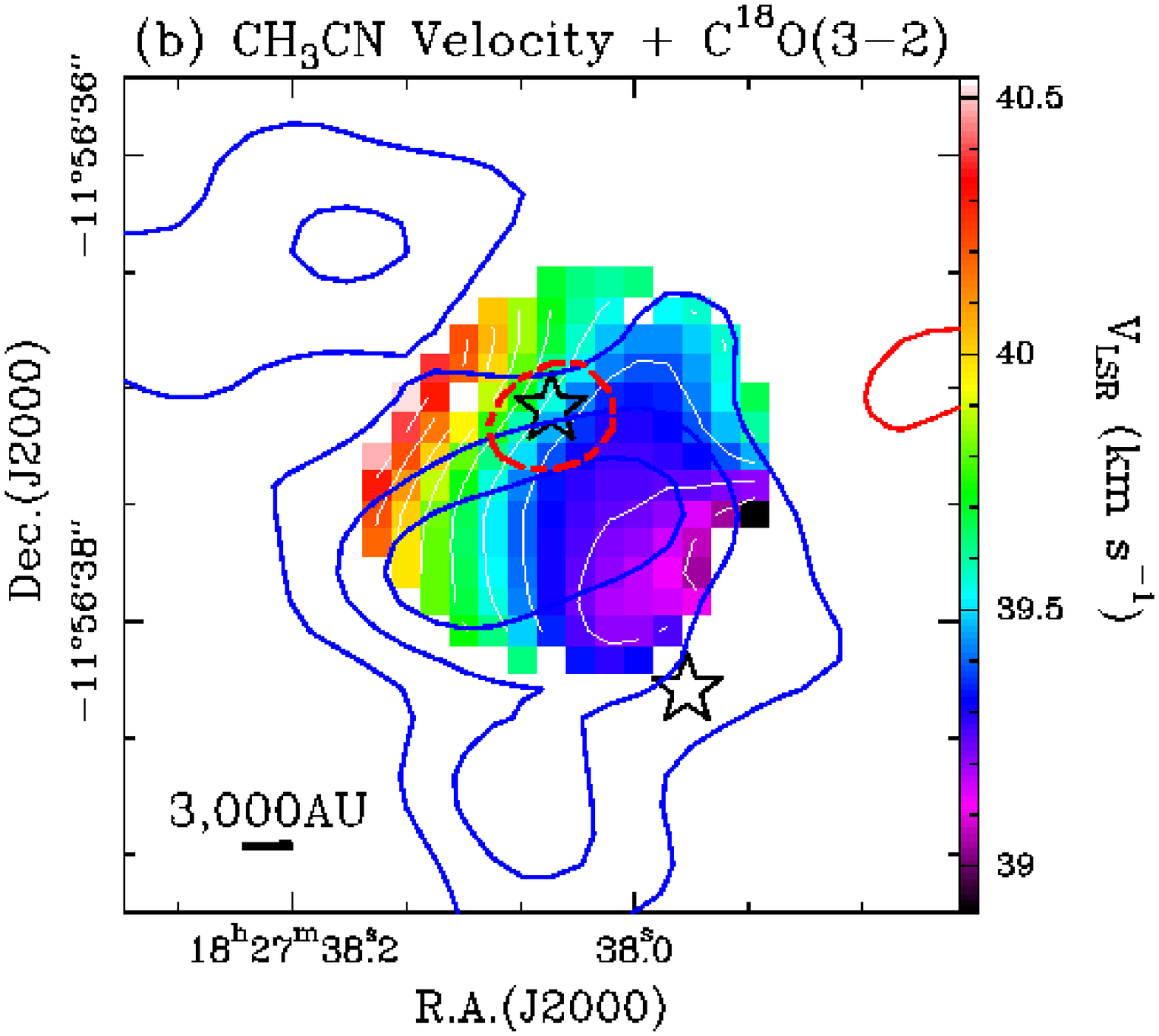}
   \caption{The color maps are the same as in Fig.~\ref{fig:velmaps},
   while the overlayed solid blue and dashed red contours represent, respectively, 
the blueshifted emission and redshifted absorption seen in the \tCO\
   (left panel) and \CeO\ (right panel).
To obtain these plots, we integrated the line
over 3 velocity channels centered on
the deepest absorption and the most intense emission, namely over the
ranges:
34.2 $\leq$\Vlsr /\kms $\leq$ 35.8 for the blueshifted \tCO\ emission,
43.8 $\leq$\Vlsr /\kms $\leq$ 45.0 for the redshifted \tCO\ absorption,
and
36.0 $\leq$\Vlsr /\kms $\leq$ 38.4 for the blueshifted \CeO\ emission,
44.0 $\leq$\Vlsr /\kms $\leq$ 46.4 for the redshifted \CeO\ absorption.
These velocity ranges are outlined by the horizontal blue and red bars
in Fig.~\ref{fig:SPhv}.
All the contours  are in steps of $\pm$2\sgm\ and start from the $\pm$3\sgm\
level.
The stars mark the peak positions of SMA1 and SMA2.
See Sect.~\ref{ss:SMA1VelStr} for details.
         \label{fig:COabsmaps}}
   \end{figure}

The velocity structure seen in \mcn, a high-density,
high-temperature tracer, is very different from that obtained from the
\tCO\ and \CeO\ (3--2) lines (see Fig.~\ref{fig:COabsmaps}).  A comparison
between the blue- and red-shifted emission in these lines and the velocity
field of the \mcn\ transition is shown in Fig.~\ref{fig:COabsmaps}. 
The most interesting result is that both the deepest
absorption and the 890~\um\ continuum
peak roughly coincide with the center of symmetry (both in space and
velocity) of the \mcn\ distribution.  This configuration is strongly
suggestive of the HMC to be both rotating about a SE--NW axis and collapsing.
The fact that no hint of infall, i.e. no redshifted (self)absorption is
detected in the \mcn\ lines can be explained by these being higher-energy
transitions than the \tCO\ and \CeO\ (3--2) lines.  Very likely, the
absorption observed in the latter arises in the outer, low-density regions of
the core where the temperature is significantly less than that of the
high-density gas traced by \mcn.
In the following, we discuss the details of this scenario and derive some
physical parameters characterizing the infalling/rotating gas.

%

\subsubsection{Infall}
\label{sss:abs}

The negative features observed in the \tCO\ and \CeO\ (3--2) spectra
are due to absorption against the innermost, optically thick part of the core.
Unlike the case of absorption at cm wavelengths, where the outer regions of a
molecular envelope absorb the free-free continuum of an embedded \hii\ region,
here we have detected absorption against the continuum emitted by the core
itself.
The problem is thus complicated by the fact that gas and dust are mixed,
so that both line and continuum photons are emitted from any point inside
the core. However, in the outer region of the core (where absorption occurs)
the density is probably low enough to decouple the line radiative transfer
from that in the continuum, whereas in the innermost region (where the bright
continuum emission comes from) the dust optical depth is large enough to
absorb all line photons. With this in mind, we can simplify the
problem assuming a spherical core made out of two regions:
an outer molecular shell enshrouding an inner, optically thick dusty nucleus.
The temperature increases outside-in, so that the nucleus is hotter than
the shell. In this configuration, the line brightness temperature measured
along the line of sight through the center of the core can be written as
\begin{equation}
 T_{\rm B} = [J_\nu(T_{\rm ex})-T_{\rm c}] (1-{\rm e}^{-\tau})
\label{eq:rt_abs}
\end{equation}
where $T_{\rm c}\simeq 31$~K (Sect.~\ref{sss:smacont}) 
is the measured continuum brightness temperature, $\tau$
is the line optical depth, and $J_\nu$ is defined as
$J_\nu(T)=h\nu/\{k\,[\exp(h\nu/kT)-1]\}$, with $h$ and $k$ Planck and
Boltzmann constants, respectively.  We have neglected the contribution of the
cosmological background temperature ($T_{\rm BG}\simeq 2.7$~K), as this is
very small at the \tCO\ and \CeO\ (3--2) line frequencies ($h\nu/k\simeq 15.9$~K,
hence $J_\nu(T_{\rm BG})\simeq 0.045$~K).\par

Equation~(\ref{eq:rt_abs}) can be written for both the \tCO\ and the
\CeO\ (3--2) lines, taking into account that the values of $T_{\rm B}$
are respectively --29~K and --15~K and $\tau$(\tCO)\,=\,6.4 $\tau$(\CeO).
From the two equations one obtains $\tau$(\tCO) $\simeq$\,4.5 and
$T_{\rm ex}\simeq7$~K. For a line width of $\sim$6~\kms, these imply a
\tCO\ column density of $\sim 1.8\times10^{17}$~cm$^{-2}$ and
an H$_2$ column density $N_{\rm H_2}\simeq10^{23}$~cm$^{-2}$, 
assuming a \tCO\ abundance of $1.5\times10^{-6}$.\par

Such a low value of \Tex\ is not consistent with the absorbing gas being
associated with the HMC and suggests that absorption could be due to a
cold layer of molecular gas located in the outer regions of the cloud.
Alternatively, the absorbing gas could cover only a fraction of the
continuum. In this case, one must introduce a filling factor $<1$ multiplying
$J_\nu(\Tex)$ in Eq.~(\ref{eq:rt_abs}) and the value of 7~K becomes a lower
limit.

Assuming that the absorbing gas is indeed associated with the HMC,
one can roughly estimate the mass accretion rate for a constant density
distribution:
\begin{equation}
\dot{M} = 4\pi\,m_{\rm H_2}\,R\,N_{\rm H_2}\,V_{\rm inf}
\end{equation}
where we take the infall speed ($V_{\rm inf}\simeq4$~\kms) equal to the
difference between the velocity of the absorption dip and the systemic
velocity (Sect.~\ref{sss:hmcsp}).  
The radius, $R$, at which absorption occurs is difficult to
estimate, but it seems clear that only the outermost layers of the core can
contribute, because the excitation temperature derived above (7~K) is much
less than the HMC temperature obtained from \mcniso\ 
(208~K; Sect.~\ref{sss:mcn_excondition}). 
Since \tCO\ is likely thermalized, 
$T_{\rm ex}$ must be very close to the gas kinetic
temperature and this implies that the radius corresponding to 7~K must be
much greater than that of the \mcniso\ emitting core. Therefore, we can only
estimate a lower limit on the mass accretion rate assuming a (minimum) value
of $R$ equal to the radius of the HMC, 0.072~pc (Table \ref{tbl:smacont}). 

We obtained $\dot{M} > 3\times10^{-3}~M_\odot\,{\rm yr}^{-1}$, 
consistent with the findings of Wu et al. (2009). 
Note, however, that this is a very conservative lower limit
as in all likelihood a temperature of 7~K pertains to gas layers located much
further than 0.072~pc from the HMC center.
Indeed, comparison with the large outflow mass loss rate 
(Sect.~\ref{sss:HtCOp}) suggests that
the actual infall rate could be much larger than the lower limit quoted above.\par

Finally, we note that such a large accretion rate is sufficient to quench an
\HII\ region even from a star as luminous as suggested by the outflow
momentum rate, i.e. $10^5$~\Lsun\ (see Sect.~\ref{sss:HtCOp}). This
corresponds to a zero-age main-sequence O7 star, which requires an infall
rate $\ge2\times10^{-5}$~\Msun\,yr$^{-1}$ to quench the
corresponding \hii\ region (see Walmsley 1995), much less than the lower
limit derived by us. It must be noted, though, that the system we are
considering is not spherically symmetric, so that one cannot exclude that a
hypercompact \hii\ region would form even under such extreme conditions.

%
   \begin{figure}
   \centering
   \includegraphics[width=7cm,angle=0]{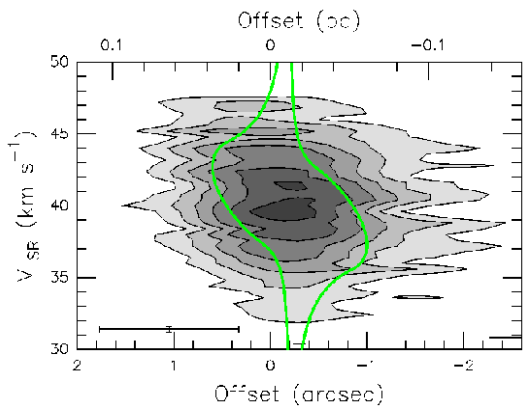} \\ 
      \caption{Position-velocity diagram of the \mcn\ $18_3-17_3$ component
toward the SMA1 core. The cut is made along a direction with P.A. $=$37\deg,
i.e. along the velocity gradient identified in Fig.~\ref{fig:velmaps}.
The contour levels range from 0.61 to 4.1 in steps of 0.6 Jy \pbeam.
The horizontal and vertical bars in the bottom left indicate
the spatial and velocity resolutions, respectively.
The solid green curves encompass the region of the plot inside
which emission is expected if the gas undergoes Keplerian rotation about and
free-fall infall onto a massive star with mass 83~$M_\odot$. 
              }
         \label{fig:PVmcnk3}
   \end{figure}

\subsubsection{Rotation}
\label{sss:VelStrCENT}
As discussed in Sect.~\ref{sss:Vmap}, we believe that
the SMA1 core is undergoing rotation about the SE--NW axis of the water
maser jet and \HtCOp\ outflow.
Is this rotation sufficient to stabilize the HMC?  We know
that infall is occurring on a larger scale than the HMC, but it is to be
understood if gravitational collapse continues inside the HMC or, instead,
the infalling material attains centrifugal equilibrium. The mass that can be
supported by rotation is $M_{\rm dyn}=R\,V_{\rm rot}^2/G\simeq83~M_\odot$, with
$G$ gravitational constant and $V_{\rm rot}$ rotational velocity at the outer
radius of the HMC, $R$. The fact that the HMC core mass obtained from the
sub-mm continuum emission ($\sim$1300~$M_\odot$) 
is much greater than $M_{\rm dyn}$ argues 
in favor of the core to be undergoing gravitational collapse.\par

To check if the idea of a an infalling and rotating HMC is consistent with
our results, we have made a position--velocity diagram of the \mcn\ (18--17)
$K$=3 line emission, along the direction of the velocity gradient (i.e.
approximately NE--SW). This is shown in Fig.~\ref{fig:PVmcnk3}, where we
have also overlayed a pattern representing the maximum and minimum velocities
expected at a given position from a rotating and collapsing core. We have
arbitrarily assumed Keplerian rotation and infall with zero velocity at
infinite distance from the HMC center. The pattern in the figure corresponds
to a central mass of 83~$M_\odot$ and an outer radius of 10100~AU.\par
Although the simple scenario depicted here is far from being unique,
the pattern is consistent with the line emission, once the limited angular
resolution is taken into account. This shows that one cannot rule out
the possibility that the infalling material is settling onto a centrifugally
supported disk in the innermost regions of the HMC.

\subsection{Nature of the SMA1 Core}
\label{ss:SMA1nature}

What is the stellar content of the HMC?  Does this consist of one (or a
few) massive stars, or is the core hosting a cluster of lower mass stars? Our
findings do not allow to draw any firm conclusion, as the molecular gas
cannot be investigated with sufficient resolution and only a loose upper
limit can be set on the HMC luminosity.  
With the advent of the next generation of large telescopes,
it will be possible to overcome these problems, but at present we can only
make speculations.\par

We have seen in Sect.~\ref{ss:evolution} that no free-free emission is
detected towards the HMC and that this may imply that no star earlier than
B0.5 is present in the core or that the star is still in a pre-UC\hii\ region
phase. If the scenario proposed in Sect.~\ref{ss:SMA1VelStr} is correct, we
are dealing with a well defined system, consisting of a massive core
undergoing infall and rotating about a water maser jet/\HtCOp\ outflow. Such
a symmetric structure and the large mass accretion rate derived suggest that
only few high-mass YSOs might be located at the center of this system, rather
than a cluster with a significant contribution from low-mass stars. This
hypothesis is supported by the comparison between the fragmentation time
scale, $t_{\rm frag}$, and the free-fall time, $t_{\rm ff}$, of the HMC. The
former can be estimated as the ratio between the core diameter and the line
width of a typical HMC tracer, e.g. \mcn:
$t_{\rm frag}=0.14~{\rm pc}/10~\kms\simeq 1.4\times 10^4$~yr. 
The latter is calculated
for a mean H$_2$ density of $\sim4\times10^7$~cm$^{-3}$, and is
$t_{\rm ff}\simeq6\times10^3$~yr. 
Because of $t_{\rm frag}\sim 2t_{\rm ff}$,
we argue that fragmentation could be partially inhibited during the collapse.\par

It is worth pointing out that the previous discussion has limited validity,
as it is mostly based on qualitative arguments and neglects the effect of the
magnetic field, which might contribute significantly to stabilize the core
against gravitational collapse. Nonetheless, we believe that our findings
lend support to the idea that the stellar content of the core could be
biased towards very young, OB-type stars.

\section{Conclusions and Future Perspectives}
\label{s:c}

We have performed deep continuum imaging from the centimeter to the
sub-millimeter regime
of the \gn\ high-mass star forming region. Our observations
have confirmed the existence of a large number of UC \hii\ regions,
as well as that of a HMC. The latter is resolved into three cores, with
one of these (here called ``the HMC'')
being $\sim$10 times more massive than the others.
We have also mapped the region
in a number of molecular lines at 3~mm and 890~$\mu$m,
most of which appear to trace the three cores. Star formation
is likely to occur not only in the HMC (SMA1), but also in
the other two (SMA2, SMA3), as witnessed by the existence of a
bipolar outflow seen in the SiO (2--1) line towards SMA2.\par

The \mcn\ (18--17) line emission reveals a velocity gradient across SMA1,
roughly perpendicular to a water maser jet and bipolar \HtCOp\ outflow
directed in the SE--NW direction. We interpret this velocity gradient as due
to rotation about the jet/outflow axis. We also confirm the existence of an
inverse P-Cygni profile in the \tCO\ (3--2) line, already detected by Wu et
al (2009) and reveal a similar profile also in the \CeO\ (3--2) line. This
redshifted absorption strongly suggests that the SMA1 core is infalling --
beside being rotating -- with a mass accretion rate $>3~10^{-3}~M_\odot\,{\rm
yr}^{-1}$.  We conclude that very young OB-type stars are likely forming
inside the HMC.\par

The study presented here is an excellent benchmark of what will
be feasible, with much better resolution and sensitivity, with
new generation instruments. Deep unbiased
surveys of selected high-mass star forming regions will permit
to improve our knowledge of the OB star formation process, identifying
newly formed stars by means of their free-free continuum emission
and investigating at the same time the structure and velocity field
of the molecular cores associated with them.

\begin{acknowledgements}
The authors are grateful to the referee, Dr. S. Curiel, and
the editor, Dr. C. M. Walmsley, for their constructive comments
to the manuscript. 
The authors sincerely acknowledge S. Takahashi, M. Momose,
L. Testi, and C. Codella for their contribution at early stage
of this study as co-authors of paper I.  
In particular, S. Takahashi and M. Momose performed the
NMA observations and calibrated most of the visibility data.
The authors also acknowledge 
M. J. Claussen and G. V. Moorsel for their extensive help in VLA observations,
especially handling the EVLA antenna data,
D. Fong for his help in calibrating the SMA data,
J. M. Carpenter and J. Lamb for their help in combining the OVRO and NMA visibilities,
and A. I. Sargent and P. T. P. Ho for their fruitful comments and encouragement.
R. S. F. is supported by a Grant-in-Aid
from the Ministry of Education, Culture, Sports, 
Science and Technology of Japan (No.\,20740113).
\end{acknowledgements}

\appendix
\section{Rotation Diagram Analysis of the SiO Lines}
\label{as:pd}

\begin{figure}[t]
   \centering
   \includegraphics[width=4.5cm,angle=0]{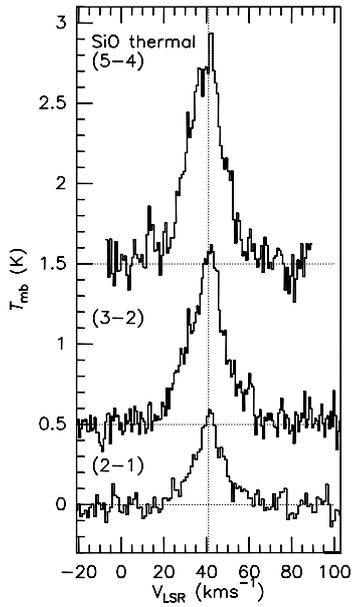}
      \caption{Spectra of thermal ($v =0$) emission of the SiO molecule 
towards \gn\ taken with the IRAM\,30\,m telescope 
(J.M. Acord personal communication) 
in main-beam brightness temperature (\Tmb) scale.
The vertical dashed-line indicates systemic velocity (\Vsys) of \Vlsr\ $=$ 41.6 \kms.
              }
         \label{fig:SPsio}
   \end{figure}

\begin{figure}
   \centering
   \includegraphics[width=6cm,angle=-90]{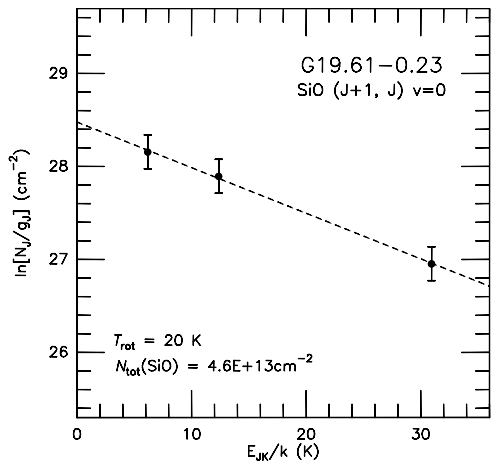}
      \caption{Rotation diagram of the SiO thermal emission shown in Fig.~\ref{fig:SPsio}. 
The dashed line is the best-fit to the data. The values of the rotation
temperature and column density thus obtained are given in the bottom left
of the figure.
}
         \label{fig:RDsio}
   \end{figure}

We estimated the excitation temperature of the SiO molecule from the (2--1),
(3--2), and (5--4) transitions (Fig.~\ref{fig:SPsio}; J. M. Acord, personal
communication) that have been observed simultaneously with IRAM 30\,m
telescope. In this way, problems due to relative gain calibration and
pointing errors are minimized. Following Acord et al. (1997), we have applied
the ``rotation diagram'' method, whose advantages and limitations are
discussed in detailed in (\cite{gl99}).  For this purpose, in
Fig.~\ref{fig:RDsio} we plot the logarithm of the column density in the lower
level of each transition, divided by the corresponding statistical weight,
against the energy of the upper level, $E_{\rm u}$. The column densities were
obtained from the line intensities under the assumption of optically thin
emission. The column densities were also corrected for the different beam
filling factors of the three lines, assuming the source to be point-like and
referring all measurements to the 28\arcsec\ beam at the frequency of the
(2--1) line.\par

The slope of the linear fit to the data, $1/T$, gives the ``rotational
temperature'' (\Trot) of the SiO molecules, which is likely an underestimate
of the kinetic temperature of the H$_2$ gas (see e.g. Acord et al.  1997).
The data are well fitted by a straight line, which indicates that the method
used is likely correct, although the SiO molecule is known to be subthermally
excited as noted above. We obtain \Trot$\simeq$20\,K and a total SiO beam
averaged column density of 4.6$~10^{13}$ \cmq.

\end{document}